\documentclass{aa}            

\usepackage{graphicx}
\usepackage{txfonts}
\begin{document} 

   \title{Scaling of  X-ray spectral properties of a black hole in the Seyfert 1 galaxy NGC~7469}

   \author{Elena Seifina,
           \inst{1,3}
Lev Titarchuk 
          \inst{2} 
\and
           Lyubov Ugolkova\inst{3}
          }

   \institute{LAPTh, Universite Savoie Mont Blanc, CNRS, B.P. 110, 
Annecy-le-Vieux F-74941, France, 
\email{seifina@lapth.cnrs.fr} \\ 
%
        \and
Dipartimento di Fisica, Universit\`a di Ferrara, Via Saragat 1, I-44122 Ferrara, Italy, 
\email{titarchuk@fe.infn.it}
 \\
        \and
Moscow State University/Sternberg Astronomical Institute, Universitetsky 
Prospect 13, Moscow, 119992, Russia;\email{seif@sai.msu.ru}
}             

   \date{Received 
        ;       accepted 
}


%


 
  \abstract
{We present our analysis of X-ray spectral properties observed  from 
the Seyfrert 1 galactic nucleus NGC~7469 using the {\it Rossi X-ray Timing Explorer (RXTE)} and 
{\it Advanced Satellite for Cosmology and Astrophysics mission (ASCA)} 
observations. 
We demonstrate 
 strong observational evidence that  NGC~7469 undergoes  
spectral transitions from the low  hard state (LHS) to the  intermediate state (IS)  during these observations.
The {\it RXTE} observations (1996--2009) show that the source was in the IS $\sim 75$\%  of the time only $\sim 25$\% of the time in the LHS. 
The spectra of NGC~7469  are well fitted by the so-{called} bulk motion Comptonization (BMC) model  for all spectral states. 
We have established the photon index ($\Gamma$) saturation level,
$\Gamma_{sat}$=2.1$\pm$0.1, in the $\Gamma$ versus  mass accretion rate, $\dot M$ 
correlation. 
This $\Gamma-\dot M$ correlation allows us to estimate the black hole (BH) mass 
in NGC~7469  to be  
$M_{BH}\geq 3 \times 10^6 M_{\odot}$ assuming the 
distance  to NGC~7469  of 70 Mpc. 
For this BH mass estimate, we use the scaling method taking Galactic BHs,  
GRO~J1655--40, Cyg~ X--1,  
and an extragalactic BH source, NGC~4051 
as reference sources. The $\Gamma$ versus $\dot M$ 
correlation revealed in NGC~7469  is similar to those in a number of 
Galactic and extragalactic BHs and it  clearly shows the correlation along with 
the strong $\Gamma$ saturation at  $\approx 2.1$. This is  robust 
observational evidence for the presence of a BH in NGC~7469. We also find that  
the  seed (disk) photon temperatures are quite low, of the order of 140$-$200 eV, which are consistent with  a high BH mass in NGC~7469 that is more than $3\times10^6$ solar masses. 
} 

\titlerunning{Evaluation of BH mass in NGC~7469}


   \maketitle
%

\section{Introduction}

Active galactic nuclei (AGNs) belong to the most luminous objects in the Universe, 
frequently {outshining} their host galaxies.  Historically, AGN sources have been 
classified using their optical and radio characteristics and were initially considered 
as a complex of different objects. It is now thought that all AGNs are the same in 
terms of a common central engine, namely, an accreting supermassive black hole (SMBH) 
located at the center of its host galaxy. Several of the observed differences in these 
sources can be explained by angle-dependence and obscuration effects. However, these properties cannot 
be enough to explain the great variety of AGN properties that have been discovered. 
It is very desirable to study AGNs in X-rays in combination with optical 
and radio emission using spectral and timing analysis.

Among such AGNs, NGC~7469 is a  famous Seyfert 1 galaxy located in{ the} Pegasus
constellation. 
In optical emission this source was discovered {more than seventy} 
years ago by Carl Seyfert
~\citep{Seyfert43}. 
X-rays from NGC~7469 were first detected by $Uhuru$ 
(Forman et al., 1978). The main feature of this galaxy 
is the activity of its central region detected in optical and X-rays 
as a compact, variable source, and the presence of an ultraviolet emission excess 
from the galaxy's central region. The galaxy NGC~7469 (Arp 298 = MCG 1-58-25 =  Mrk 1514) 
is classified as a spiral SBa galaxy slightly inclined to the line of sight. 
The distance to NGC~7469 is estimated to be in a wide range (from 50 to 117~Mpc) obtained by different methods: 
the AGN time lag (47 -- 117 Mpc; \cite{Peterson04,Peterson14}), SN~Ia (66~Mpc; \cite{Ganeshalingam13}], 
Tully-Fisher (40 -- 61 Mpc; \cite{Bottinelli84,Theureau07}), and others. 

We take the rest frame of NGC~7469 to be at  a redshift of z=0.016268, based on the 21 cm line 
\citep{Spirandob05,Peterson04}, and thus luminosity distance to be 70 Mpc. 
Analyses of X-ray $Chandra$ and XMM-$Newton$ grating spectra of NGC~7469 \citep{Scott05} 
provide 
the total absorbing column density in the X-rays of order 3$\times$10$^{21}$ cm$^{-2}$.
For the neutral absorption in our Galaxy, we use a column density of 6$\times$10$^{20}$ cm$^{-2}$ 
that includes both H1 (4.34$\times$10$^{20}$ cm$^{-2}$) and H$_2$ (5.75$\times$10$^{19}$ cm$^{-2}$, \cite{Wakker11}), 
for a total of $\sim 3\times 10^{21}$ cm$^{-2}$ \citep{Behar17}.
Around the central part of this galaxy,  two ring areas of star formation are observed. 
The central region of NGC~7469 exhibits variability in 
the X-ray, ultraviolet, optical, and infrared bands as well as in spectral lines. 
However, NGC~7469 is a weak radio  source  as are most of all known Seyfert galaxies.

\begin{figure*}
 \centering
    \includegraphics[width=15cm]{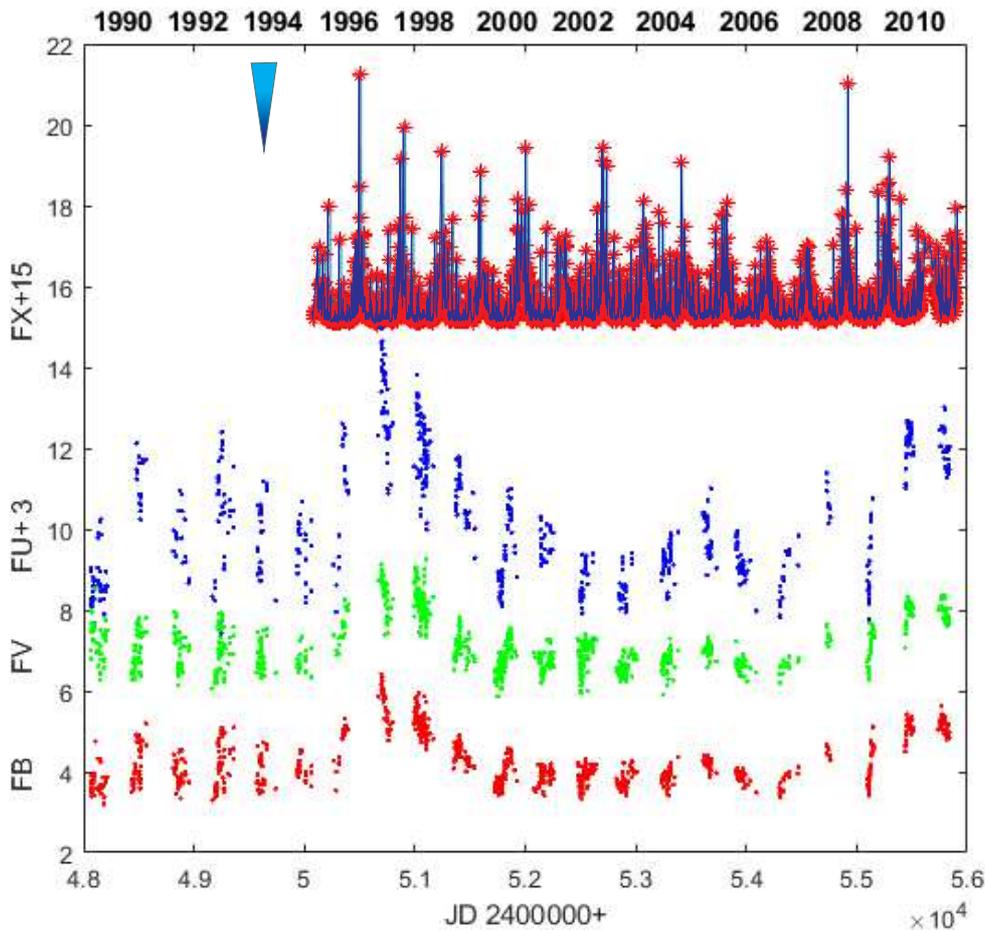}
      \caption{
X-ray light curve of NGC~7469 in the 2$-$12 keV energy range (1996 -- 2011) is presented on the top.  
{The bright blue triangle (arrow) indicates the {\it ASCA} observation MJD.} 
The rate axis for the upper light curve (red stars) is related to the {\it RTXE}/ASM count rate increased by 15 units for clarity. 
In the three bottom panels the optical light curves (in $10^{15}$ erg/cm$^2$/s/$\dot  A$) of NGC~7469 in the U (blue), V (green), and B (red) filters 
 during 1995--2012 are presented. For clarity, the flux for the U-filter is increased by three units. 
}
   \label{lc_0}
 \end{figure*}

NGC~7469 is well studied in optical {emission} 
(see, e.g., \citet{Doroshenko89}, 
\citet{Merkulova00}, 
\citet{Sergeev05}) 
using the facilities of the Crimean Astrophysical Observatory and the Crimean 
Laboratory of the Sternberg Astronomical Institute (U, B, V  light curves are presented in Fig.~\ref{lc_0}). It is worth noting  that the AGN Watch 
international project organized in the mid 1990s~(see \cite{Alloin94}) 
obtained {numerous} 
results on
the optical variabilities of NGC~7469 (see   \cite{Peterson97,Ulrich97,O_Brien+Leighly98})  
 and 
on ultra-rapid variability \citep{Dultzin-Hacyan92,Dultzin-Hacyan93}. 
Results of the Hubble Space Telescope and the {\it Multicolor active Galactic Nuclei Monitoring} (MAGNUM) 
telescope (2001 -- 2003) observations of NGC~7469 
are presented in  Suganuma et al. (2006). 
 Analysis of MAGNUM light curves indicates the presence of short-term variability, 
 from several  days to several weeks. Optical observations of  NGC~7469, particularly  
 observations performed in the Maidanak observatory (Artamonov et al. 2010), 
 indicate  long-term (eight -- nine years)  variability of the nucleus with outbursts 
 of a relatively long rise (one -- two years), peak  ($\sim$ one -- two years), and long 
  decay (three years).
  
Several giant outbursts in NGC~7469 are known to have been  observed in UBVRI bands  
(see bottom panels in Fig.~\ref{lc_0}) with maxima in 
1993 \citep{Lyutyi95,Lyuty05,Dorosh98}, 1997--1998, 
 2005 \citep{Doroshenko10,Artamonov10,Ugolkova11},  
 and 2012 \citep{Ugolkova17,Shapovalova17}. 
Unfortunately, X-ray pointed observations with  the {\it Rossi X-ray Timing Explorer (RXTE)} 
cover only 
partial phases: a rise (1996) of the 1994--2001  outburst, a decay in 2003, and a rise  in 2009 (see \cite{Ugolkova17}),  
while soft X-ray monitoring observations 
(ASM, 2 -- 10 keV)\footnote{HEASARC, http://xte.mit.edu/ASM\_lc.html.}
cover continuously the whole (1996--2011)  interval (see top panel in Fig.~
\ref{lc_0}). 
It is worth noting  that the soft X-ray emission from NGC~7469 demonstrates a more {complex} 
pattern than that in the optical band (compare top and bottom panels in Fig.~\ref{lc_0}). 
Generally, the source shows numerous short-term X-ray outbursts (on the order of  a few days) and sparse long-term optical 
outbursts (six-- eight years), modified by rapid variability (for more details see Sect.~3.1 ). 
A stellar disruption was suggested as an explanation for the great outburst that occurred  in 1997--1998  
~\citep{rees88}. 
In the present paper we  reanalyzed X-ray 
observations 
and matched them with  optical ones in order to diagnose  the source activity states based on X-ray/optical 
light curve pattern. 

It is important to emphasize  that 
the optical$-$X-ray connections observed in NGC~7469 are very different for various epoches.
To explain these differences, 
Gaskell (2006, 2007) suggested a model using X-ray anisotropy 
in combination with 
an effect  of the 
dust  surrounding the  accretion disk, which can cause 
the lags between variations in X-ray and optical bands. 
The physics of  processes near the nucleus of NGC~7469 are 
also  discussed in detail by Doroshenko et al. (2010), Chesnok et al. (2009), and 
 Ugolkova et al. (2017).  Using a comparison of the X-ray and optical properties of NGC~7469, 
 these authors  concluded that the  direct Compton X-ray reprocessing cannot provide the whole observed 
 optical luminosity of NGC~7469. 
  Other processes, such as thermal radiation from the accretion disk, the star 
  formation processes, and inverse Compton scattering by hot coronal electrons  
  may also contribute to the optical emission. 
  
The mass of the supermassive black hole  (SMBH) centered in NGC~7469 is estimated {over} a wide range of 
 $M = (1 - 6)\times 10^7$ M$_{\odot}$ \citep{Peterson04,Peterson14,Shapovalova17} by 
 using the reverberation mapping method. 
It is desirable to use an independent black hole (BH) identification 
for its central object as well as the mass estimate of its  BH by an alternative to the abovementioned method. 
A  method of  BH mass determination was developed by Shaposhnikov \& Titarchuk (2009), hereafter ST09, 
using a correlation scaling between X-ray spectral and timing (or mass accretion rate) properties observed 
from many Galactic BH binaries during BH 
state transitions. 

We apply the ST09 method to the {\it Advanced Satellite for Cosmology and Astrophysics mission (ASCA)} 
and {\it RXTE} data of NGC~7469. 
Previously, many X-ray properties of NGC~7469 were  analyzed using the {\it European X-ray Observatory Satellite} (EXOSAT), 
GINGA,  
and {\it ASCA} data,
which {enabled the detection} and study {of} the soft excess of source X-ray spectra 
(Barr 1986; Turner et al., 1991; Turner et al., 1993). The GINGA observations revealed evidence 
for 6.4 keV iron florescence lines 
and flattening of the spectral shape above 10 keV (Piro et al., 1990). Spectral fits using 
a Comptonization model (Matt et al., 1991; George \& Fabian, 1991) yielded a power-law photon index of $\Gamma\sim 2$ (Piro et al., 1990; Nandra 1991).
{ Scott et al. (2012) investigated simultaneous X-ray, far-ultraviolet, and near-ultraviolet spectra of  NGC 7469 
using the Chandra X-Ray Observatory, the Far Ultraviolet Spectroscopic Explorer, and the Space Telescope Imaging 
Spectrograph on the Hubble Space Telescope. In particular, they detected   O VIII emission as well as a prominent 
Fe K$_{\alpha}$ emission line and absorption due to H-like and He-like neon, 
and  magnesium in the $Chandra$ spectrum
}

In this paper we present an analysis of  the available  $ASCA$ and $RXTE$ observations 
of NGC~7469 to reexamine previous 
conclusions on the BH nature of  NGC~7469 as well as to find further indications 
of a supermassive BH in NGC~7469.  
In Sect. 2 we present the list of observations used in our data analysis while 
in Sect. 3 we provide  details of the X-ray spectral analysis.  We discuss an evolution of 
the X-ray spectral properties during the spectral 
state  transitions 
and present the results of the scaling analysis to estimate a BH mass for NGC~7469 in Sect. 4. 
We  make our final conclusions on  the results  in Sect. 5. 

%
%

%
%

  \begin{figure*}
 \centering
    \includegraphics[width=17.9cm]{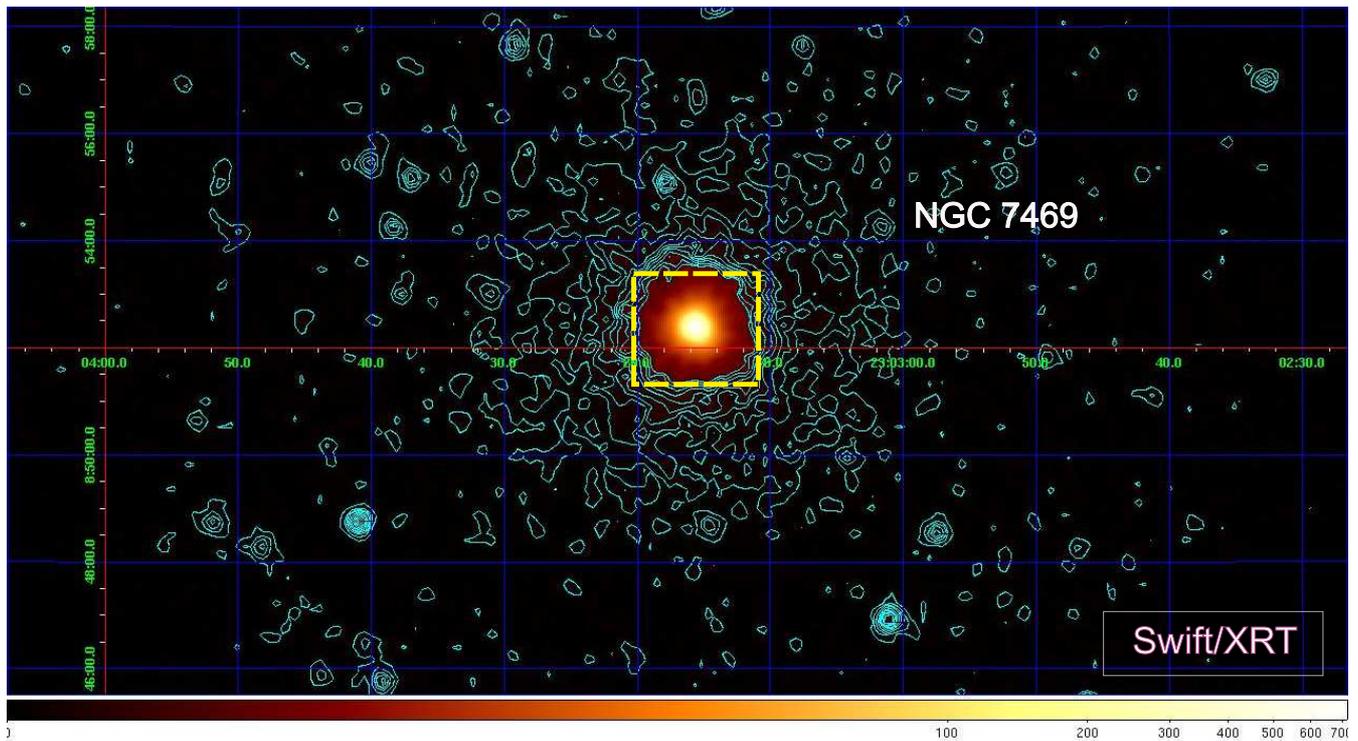}
      \caption{$Swift$/XRT (0.3$-$10 keV) image of NGC~7469 field of view. 
 Contour levels demonstrate the absence of the X-ray jet-like (elongated) structure  and the minimal contamination by other point sources and diffuse emission 
around NGC~7469. 
The image segment selected by the yellow dashed-line box is also shown in Fig.~\ref{imageb} with more details. 
}
      \label{imagea}
 \end{figure*}

%
%

  \begin{figure*}
 \centering
    \includegraphics[width=15cm]{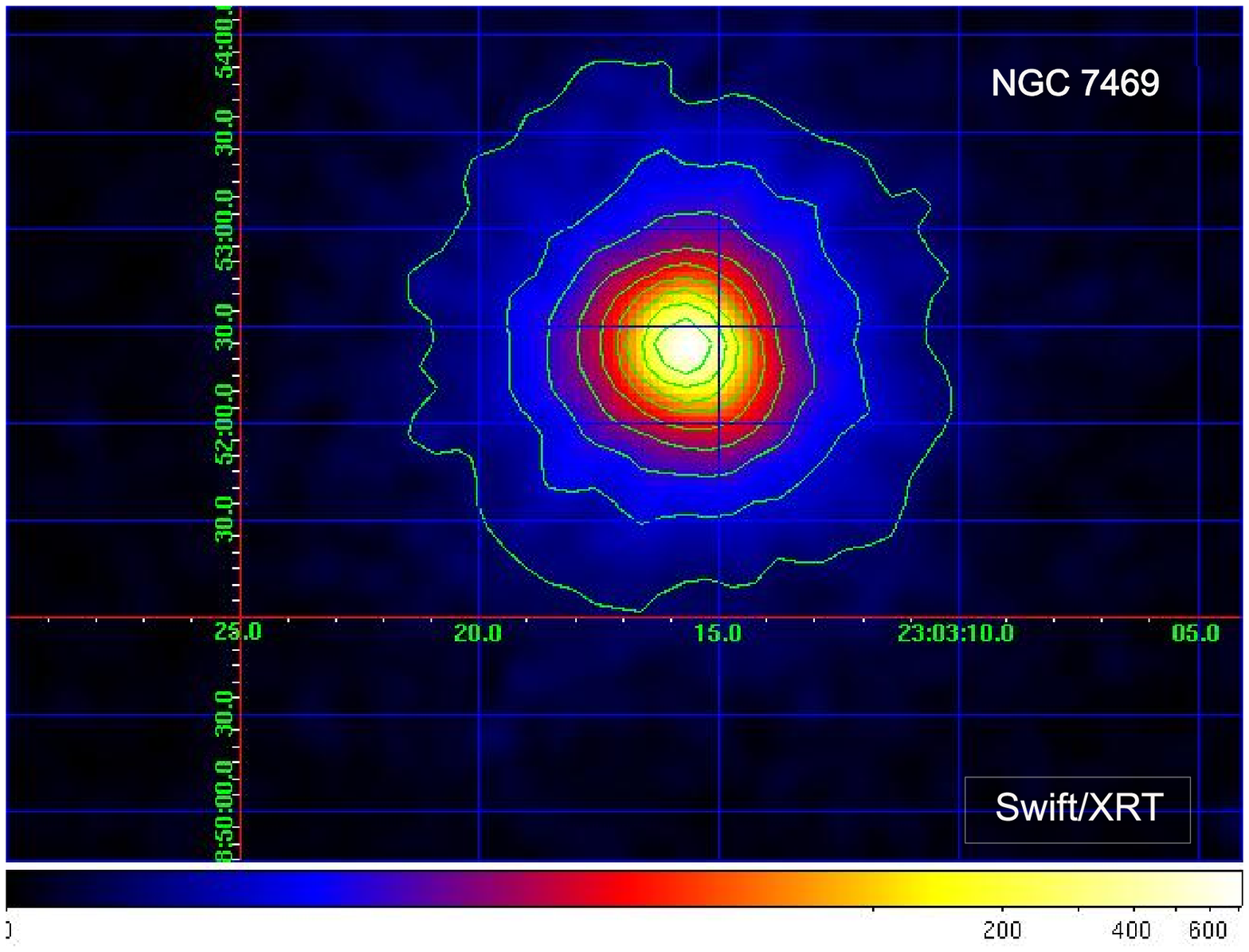}
      \caption{
Adaptively smoothed $Swift$/XRT image  of the soft X--ray emission (0.3--10 keV) of NGC~7469 field, 
of which localization is indicated by the selected  
 yellow dashed-line 
box in Fig.~\ref{imagea}.
Contours correspond to nine logarithmic intervals in the range of 0.003--5\% with respect to the brightest pixel, and 
 contour levels demonstrate the minimal contamination by other point sources and diffuse emission. 
}
\label{imageb}
\end{figure*}

\section{Observations and data reduction \label{data}}


Along with the long-term {\it RXTE} observations in 1996, 2003, 2005--2006, and 2009, 
which we describe 
in Sect.~\ref{rxte data}, NGC~7469 was observed by 
 $ASCA$ (1993, 1994, see Sect.~\ref{asca data}). 
We extracted these data from the {\it High Energy Astrophysics Science Archive Research Center} (HEASARC) 
archives and found that these data  cover 
a wide range of X-ray luminosities.  We  should also recognize that the well-exposed {\it ASCA} 
data are  preferable for the determination of   low-energy photoelectric absorption.

%

%
%

 \begin{figure*}
 \centering
 \includegraphics[width=12cm]{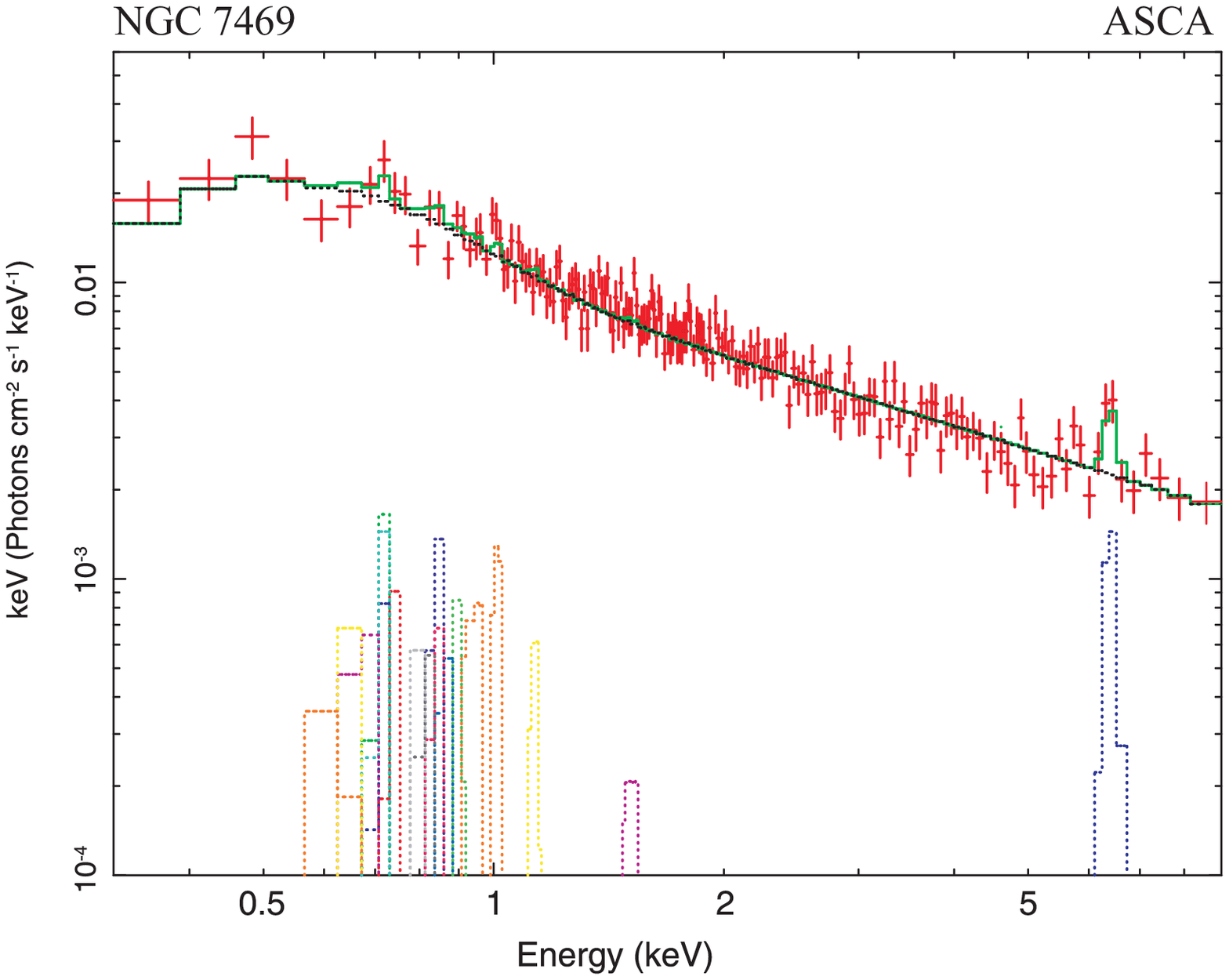}
   \caption{Best-fit {\it ASCA} spectrum of NGC~7469 for the IS 
   (ID=71028000) using  phabs*(bmc+N*Gaussian)*edge model 
   ($\chi^2_{red}=$ 1.09 (137 dof)). The best-fit parameters are 
   $\Gamma=$ 1.8$\pm$0.1, $T_s=$ 140$\pm$30 eV, and $E_{line1}=$ 6.4$\pm$0.1 keV 
   (see more details in Tables~\ref{tab:par_asca} and~\ref{tab:par_asca_lines}). 
   The data are denoted by red {crosses}, while the spectral model presented 
   by components is shown using  green and blue lines for BMC, and Fe K$_{\alpha}$ 
   Gaussian components, respectively. The narrow line components below 5 keV 
   are presented by different color lines.
}
\label{asca_interm_spectrum}
\end{figure*}

%

\begin{figure*} 
\begin{center}
     \includegraphics[width=14cm]{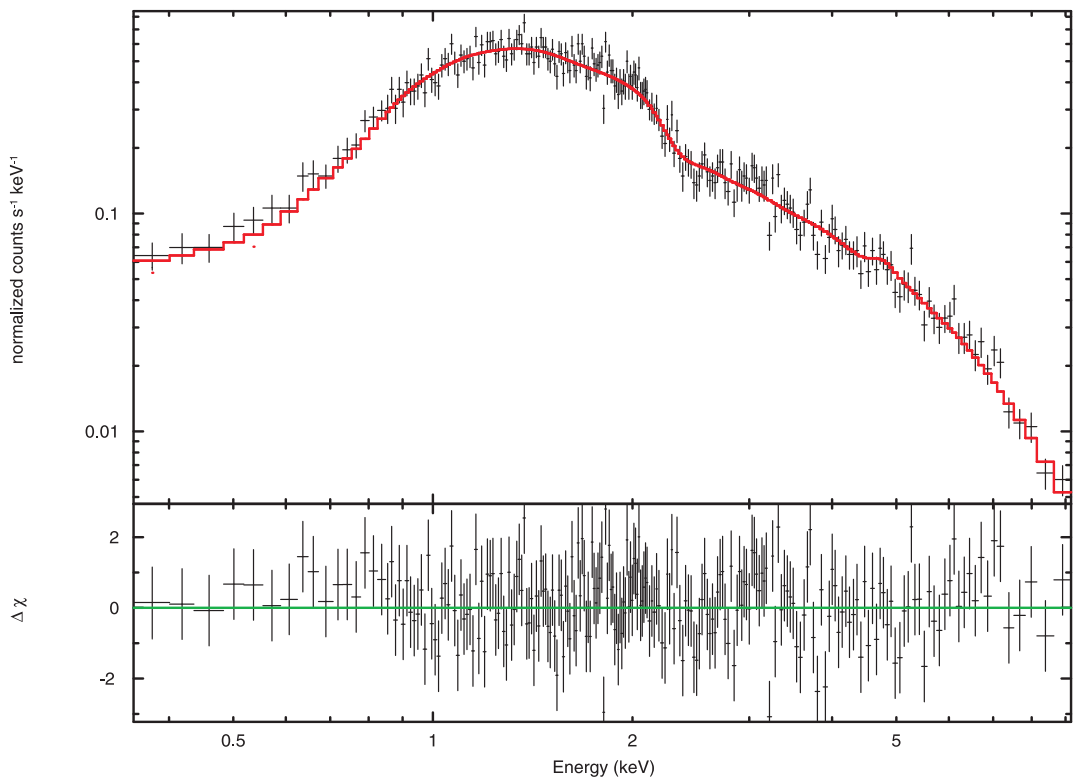}
\caption{Best-fit modeling of the spectrum of NGC~7469 (top panel) with $\Delta\chi$ residuals (bottom panel). 
Data are taken from $ASCA$ observations (ID=15030000) carried out on 1994 June 26--27, when the source 
was in the low/hard state.
}
\label{sp_ASCA_cnt}
\end{center}
\end{figure*}



\begin{table*}
 \caption{List of $ASCA$  observations of  NGC~7469 used in our analysis.}              
 \label{tab:list_ASCA}      
 \centering                                      
 \begin{tabular}{l l l l l c c}          
 \hline\hline                        
Number of set  &  Obs. ID        & Start time (UT)  && End time (UT) &MJD interval \\    
 \hline                                   
A1  &    71028000       & 1993 Nov 24  05:28:38 & & 1993 Nov  26 07:12:04 & 49315.2 -- 49317.3$^1$ &\\
A2  &    71028010       & 1993 Dec 2   05:48:22 & & 1993 Dec  3  01:22:04 & 49323.3 -- 49324.1 &\\
A3  &    71028030       & 1993 Nov 26  09:21:04 & & 1993 Nov 26  19:23:14 & 49317.5 -- 49317.8 &\\
A4  &    15030000       & 1994 June 26 23:02:40 & & 1994 June 27 10:10:18 & 49529.9 -- 49530.9 &\\
A5  &    18931126       & 1993 Nov 26  07:39:12 & & 1993 Nov  26 09:13:22 & 49317.3 -- 49317.4 &\\
 \hline                                             
 \end{tabular}
 \tablebib{
(1) Guainazzi et al. 1994.
}
 \end{table*}

%
%


\subsection{\it ASCA data \label{asca data}}

$ASCA$ observed NGC~7469 on  November 24--26, 1993. 
Table~\ref{tab:list_ASCA} summarizes the start time, end time, and the MJD interval  for each of these observations.
One can see a description of the {\it ASCA} experiment in Tanaka, Inoue, \& Holt (1994). 
The solid imaging spectrometers (SIS)  operated in Faint CCD-2 mode.
The {\it ASCA} data were screened using the ftool  ascascreen and the standard screening criteria. The
spectrum for the source was extracted using spatial regions with a diameter of 3${\tt '}$ (for {Solid-State imaging 
Spectrometer} (SISs) and 4${\tt '}$ (for 
{\it Gas Imaging spectrometer} (GIS) 
centered on the nominal position of NGC~7469 ($\alpha=23^{h}03^{m}15^s.75$, $\delta=+08^{\circ} 52{\tt '} 25{\tt ''}.9$, J2000.0), 
while background was extracted from source-free regions of comparable size away from the source. The spectrum data 
were rebinned to provide at least 20 counts per spectral bin to validate the
use of the $\chi^2$-statistic. The SIS and GIS data were fitted using {\tt
XSPEC} in the energy ranges of 0.6 -- 10 keV and 0.8 -- 10 keV, where the spectral responses
are best known. 

The source count rate was variable by a factor of 40\%. 
We fit the spectral data using a number of models (see Sec.~3.2.1) but the best-fitting is obtained with 
the sum of XSPEC BMC (see Titarchuk et al. 1997) and Gaussian models{ while taking into account absorption} at the low energies 
by neutral gas with solar composition. 
Using this model the amplitude of X-ray flux variability is up to a factor of two. 
The results of the fits are given in Table~\ref{tab:par_asca}. 

%
%

\begin{table*}
 \caption{Best-fit parameters  of the $ASCA$ spectra
 of NGC~7469 in the 0.45$-$10~keV 
 range using the following five 
models$^\dagger$: phabs*power, phabs*bbody, phabs*(bbody+power), phabs*bmc, and  phabs*(bmc+Gaussian). 
}
    \label{tab:par_asca}
 \centering 
 \begin{tabular}{lllllll}
 \hline\hline
     & Parameter & 71028000 & 71028010 & 71028030 & 15030000 & 18931126 \\
 \hline                                             
Model &      &        &        &       &      &     \\
 \hline                                             
phabs       & N$_H$ ($\times 10^{21}$ cm$^{-2}$) & 2.7$\pm$0.5 & 2.5$\pm$0.9 & 1.8$\pm$0.09  & 2.6$\pm$0.2 & 2.9$\pm$0.6 \\
Power law   & $\Gamma_{pow}$    & 2.34$\pm$0.02 & 2.29$\pm$0.02 & 2.28$\pm$0.03 & 2.31$\pm$0.02 & 2.21$\pm$0.04  \\
            & N$_{pow}^{\dagger\dagger}$ & 0.019$\pm$0.003 & 0.019$\pm$0.001 & 0.015$\pm$0.001 & 0.036$\pm$0.001 & 0.017$\pm$0.001 \\
      \hline
           & $\chi^2$ {\footnotesize (d.o.f.)} & 2.17 (214)     & 2.55 (335)     & 1.85 (183) & 1.56 (129)   & 0.92 (161) \\
      \hline
phabs      & N$_H$ ($\times 10^{21}$ cm$^{-2}$) & 3.7$\pm$0.2 & 3.5$\pm$0.3 & 4.7$\pm$0.5  & 5.8$\pm$0.3 & 2.9$\pm$0.08 \\
Blackbody      & T$_{BB}$  (keV)   & 0.54$\pm$0.01   & 0.36$\pm$0.01  & 0.95$\pm$0.01    & 0.89$\pm$0.01  & 0.26$\pm$0.01 \\
           & N$_{BB}^{\dagger\dagger}$ & 5.8$\pm$0.4 & 4.1$\pm$0.1 & 12.3$\pm$0.4 & 5.7$\pm$0.2 & 27.9$\pm$0.7 \\
      \hline
           & $\chi^2$ {\footnotesize (d.o.f.)} &  8.61 (214) & 14.1 (335) & 2.63 (183)& 4.08 (129)& 15.76 (161) \\
      \hline
phabs      & N$_H$ ($\times 10^{21}$ cm$^{-2}$) & 2.7$\pm$0.3 & 2.7$\pm$0.2 & 2.7$\pm$0.6  & 3.2$\pm$0.4 & 3.0$\pm$0.1 \\
Blackbody      & T$_{BB}$ (keV)  & 1.8$\pm$0.1 & 1.87$\pm$0.07  & 2.1$\pm$0.2  & 1.72$\pm$0.09  & 1.8$\pm$0.2     \\
           & N$_{BB}^{\dagger\dagger}$ & 3.3$\pm$0.4  & 3.4$\pm$0.3   & 2.8$\pm$0.4   & 3.5$\pm$0.4 & 3.0$\pm$0.7  \\
Power law  & $\Gamma_{pow}$    & 3.16$\pm$0.07 & 3.07$\pm$0.05 & 2.98$\pm$0.08  & 3.02$\pm$0.07 & 2.9$\pm$0.1 \\
           & N$_{pow}^{\dagger\dagger}$ & 0.03$\pm$0.06& 0.024$\pm$0.04 & 0.019$\pm$0.001  & 0.04$\pm$0.01 & 0.02$\pm$0.01 \\
      \hline
           & $\chi^2$ {\footnotesize (d.o.f.)}& 1.12 (212) & 1.11 (333) & 1.01 (181) & 0.96 (127)& 0.62 (159) \\
      \hline
      \hline
phabs      & N$_H$ ($\times 10^{21}$ cm$^{-2}$) & 3.5$\pm$0.6 & 3.1$\pm$0.6 & 2.9$\pm$0.1  & 2.8$\pm$0.3 & 3.0$\pm$0.1 \\
bmc        & $\Gamma_{bmc}$    & 1.8$\pm$0.1 & 1.86$\pm$0.04& 1.74$\pm$0.07   & 1.81$\pm$0.06    & 1.64$\pm$0.09  \\
           & T$_{s}$   (eV)    & 140$\pm$30      & 160$\pm$20   & 200$\pm$40    & 220$\pm$30     & 210$\pm$20   \\
           & logA$$            & 0.33$\pm$0.07  & 0.36$\pm$0.02& 0.34$\pm$0.05   & 0.32$\pm$0.02  & 0.36$\pm$0.05 \\
           & N$_{bmc}^{\dagger\dagger}$ & 7.56$\pm$0.04 & 7.2$\pm$0.1 & 6.1$\pm$0.1  & 6.9$\pm$0.1  & 6.58$\pm$0.05 \\
      \hline
phabs*bmc & $\chi^2$ {\footnotesize (d.o.f.)}& 1.08 (212) & 1.19 (333) & 1.09 (181)& 1.12 (227)& 0.89 (159) \\
      \hline
Gaussian   & E$_{line}$ (keV)    & 6.4$\pm$0.1 & 6.3$\pm$0.1   & 6.4$\pm$0.2 & 6.41$\pm$0.03  & 6.38$\pm$0.09  \\
           & $\sigma_{line}$ (eV) & 200$\pm$10  & 150$\pm$60    & 170$\pm$90  & 110$\pm$30     & 160$\pm$30   \\
           & N$_{line}$          & 8.9$\pm$0.3 & 3.7$\pm$0.5   & 5.3$\pm$0.1 & 1.6$\pm$0.2    & 9.4$\pm$0.5 \\
      \hline
phabs*(bmc+Gaussian) & $\chi^2$ {\footnotesize (d.o.f.)}& 1.01 (209) & 1.12 (330) & 1.08 (178)& 1.07 (224)& 0.95 (156) \\
      \hline
 \hline                                             
 \end{tabular}
\tablefoot{ 
$^\dagger$     Errors are given at the 90\% confidence level. 
$^{\dagger\dagger}$ Normalization parameters of blackbody and bmc components are in units of $L^{soft}_{34}/d^2_{10}$ erg s$^{-1}$ kpc$^{-2}$, 
where $L^{soft}_{34}$ is  soft photon luminosity in units of $10^{34}$ erg s$^{-1}$, $d_{10}$ is the distance to the 
source in units of 10 kpc, while power-law and Gaussian components are in units of 
keV$^{-1}$ cm$^{-2}$ s$^{-1}$ at 1 keV and 10$^{-5}\times$ total photons cm$^{-2}$ s$^{-1}$ at 1 keV, respectively. 
$
T_{BB}$ and $T_{s}$ are the temperatures of 
the blackbody and seed photon components, respectively (in keV and eV). 
$
\Gamma_{pow}$ and $\Gamma_{bmc}$ are the indices of the { power law} 
and bmc, respectively. 
}
 \end{table*}


\subsubsection{\it RXTE data \label{rxte data}}



\begin{table*}
 \caption{List of {\it RXTE} observations of NGC~7469.}
 \label{tab:list_RXTE}
 \centering 
 \begin{tabular}{l c c c }
 \hline\hline                        
Number of set  & Dates, MJD & RXTE Proposal ID&  Dates UT \\
 \hline    
R1  &    50185--50251      & 10293, 10315$^1$    & Apr. 12 -- June 27, 1996       \\
R2  &    52737--53001      & 80152$^1$           & Apr. 8 -- Dec. 28, 2003        \\
R3  &    53373--54097      & 90154, 91138, 92108$^1$    & Apr. 10, 2005 -- Dec. 29, 2006  \\
R4  &    54836--55194      & 94144$^1$           & Jan. 5, 2009 -- Dec. 29, 2009   \\
 \hline                                             
 \end{tabular}
   \label{tab_rxte}
 \tablebib{
(1) Rivers et al. 2013. 
}
 \end{table*}

We  also analyzed 263 {\it RXTE} observations  made between April 1996 and 
December 2009 
using the public archive.
Standard tasks of the LHEASOFT/FTOOLS 5.3 software package were 
used for data processing. For spectral analysis we 
used PCA {\it Standard 2} mode data, collected in the 3--23~keV energy range, applying 
{\it Proportional Conter Array} (PCA) 
response calibration (ftool pcarmf v11.7).
{
The fitting was carried out applying the standard XSPEC v 12.6.0 fitting package. 
}
The standard dead time correction procedure was applied to the data. 
 The data from HEXTE detectors have  been also used in order to construct the broad-band spectra.
 The spectral analysis of  the data  in the 19--50~keV energy range was implemented  
to account for the uncertainties in the HEXTE response and 
background determination.
We  subtracted a background corrected  in  off-source observations. 
 The data are available through the {\it Goddard Space Flifgt Center} (GSFC) 
public archive.\footnote{http://heasarc.gsfc.nasa.gov.} 
 We modeled the {\it RXTE}  spectra using XSPEC astrophysical fitting software and
we implemented a systematic uncertainty  of 0.5\%  to all  analyzed spectra. 
In  Table~\ref{tab:list_RXTE} we listed the  groups  of {\it RXTE} observations tracing 
through 
the source evolution during different states. 

\section{Results \label{results}}


\subsection{Images and light curves \label{lc}}

{To avoid a possible contamination from nearby sources we used  the $Swift$ data 
\citep{Gehrels04} and made a visual inspection 
of the obtained image  (after Gaussian filter smoothing). 
$Swift$ data processing is  described in our previous paper 
(e.g., Titarchuk \& Seifina, 2015). The
{\it Swift}/XRT (0.3 -- 10 keV) image of NGC~7469's field of view is presented in 
Fig.~\ref{imagea}.  The image segment selected by the yellow dashed-line box is also shown 
in Fig.~\ref{imageb} with more details. Here, contours correspond to nine logarithmic 
intervals in the range of 0.003--5\% with respect to the brightest pixel.
The image obtained during observations of NGC~7469 between April 27, 2004 and October 7, 2008 
(with exposure time of 66 ks) is presented in Fig. \ref{imageb}. 
We point out this  region as selected by  the yellow dashed-line box in the larger 
(2$'\times 2'$) {\it Swift} image in Fig.~\ref{imagea}. 
Contour levels demonstrate the absence of the X-ray jet-like (elongated) structure  and the minimal contamination by other point sources and diffuse emission around NGC~7469. This image (Fig.~\ref{imageb}) shows that the X-ray emission region is extended  up to ~5 kpc. 
} 

Before proceeding with details of the spectral fitting, we study the long-term behavior of NGC~7469, 
in particular its activity patterns. 
We compare  the one-day average X-ray and optical light curves received in the period from 1996 to 2012. 
We present a long-term daily  light curve of NGC~7469 detected by the {\it Assistant Store Manager} (ASM) 
onboard the {\it RXTE} 
over the total life time of the mission   (see top panel in Fig.~\ref{lc_0}). 
{ The bright blue triangle (arrow) indicates the {\it ASCA} observation MJD.} 
{Red} points show  the source signal 
and {the blue} line indicates the mean count rate level 
. The  ASM monitoring observations are 
distributed more densely over time than that for an optical band. However, 
optical/UBV light curves of NGC~7469 show 
a slow (long-term) variability ($\sim$ 6 
approximately six-year. 
Below, in the three bottom panels of Fig.~\ref{lc_0}, the optical light curves (in $10^{-15}$ erg/cm$^2$/s/$\AA$)  
in the U (blue), V (green), B (red) filters are related to the  1995$-$
2012 observations.  
We  found  at least one strong global outburst of NGC~7469 (1996--1997) peaked at 
around MJD=55900, 
and intervals of moderate variability with local bursts (2003--2005) and a tendency for an 
outburst rise towards 2012. This relatively slow variability is clearly seen in an optical/UBV range 
and is less evident in soft X-rays (2--12 keV, ASM). 

As seen from Fig.~\ref{lc_0}, the slow variability of NGC~7469 emission is superposed with small flares at timescales 
larger than half a day (so called mild variability). 
The modulation depth in the soft X-ray and optical bands is
typically 35\% and 5\%, respectively. 
The optical light curves (UBV) demonstrate the 
highest global maximum peaking in the 1996--1997 period (see three bottom panels in Fig.~\ref{lc_0}), while 
the X-ray light curve shows more significant variability with frequent outbursts (top panel in Fig.~\ref{lc_0}). 

It is worth noting that the peaks of X-ray outbursts in the ASM (2 -- 12 keV) light curve are 
characterized by variable amplitudes, but at the maxima of the  X-ray peaks they are closely related to 
the optical maximum (1996--1997). This fact can indicate  activity of NGC~7469 in 
a wide photon energy range, in spite of the fact that the optical and  X-rays presumably originated in geometrically different areas. 
The slow variability by a factor of ten has been seen in the 1996--2012
observations by ASM/{\it RXTE}. The same kind of changes of the
flux were also observed in the earlier observations by \cite{Ugolkova11}. 
One can relate slow and mild variabilities of 
NGC~7469 to slow and mild changes of the mass accretion rate. 
Therefore, it is interesting how the slow and mild variabilities affect the spectral properties of NGC~7469.

\subsection{Spectral analysis \label{spectral analysis}}

Different spectral models were used  in order to test them  for all available data  
for NGC~7469. We want to establish  the low/hard  state (LHS) and the intermediate state evolutions 
using  these spectral models.
We investigate  the  $ASCA$ and $RXTE$ spectra 
to test  the following XSPEC spectral models: 
power law, blackbody
, {\it Bulk Motion Comptonization} (BMC)
, and their possible combinations modified by absorption and Gaussian models. 

\subsubsection{Choice of the spectral model\label{model choice}}


Our result is  in agreement with previous results from {\it ASCA}. In particular, when the source was in 
the intermediate state, Guainazzi et al. (1994) found a photon index of $\Gamma=2.003\pm 0.008$ (for A1 observation), while our result differs from previous results using  the EXOSAT and GINGA data. The EXOSAT data fits found  $\Gamma=1.78\pm 0.07$ (see Turner \& Pounds 1987), 
which is close to that for  the GINGA ones, $\Gamma=1.83\pm 0.01$ (Piro et al., 1990), when the source presumably  was in the low-hard state.
However, these results for the {\it ASCA},  EXOSAT and GINGA data  suggest a spectral variability, 
which is better seen in broader bandpass than those in the {\it ASCA}, EXOSAT and GINGA energy ranges.

%
%
\begin{figure*}
\begin{center}
 \includegraphics[width=10cm]{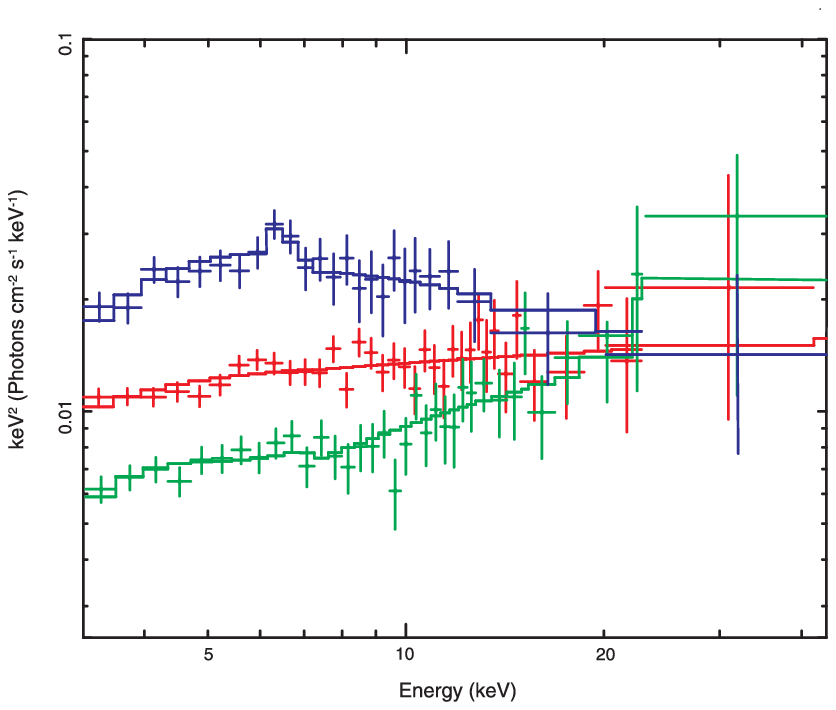} 
\end{center}
  \caption{
Three representative $EF_E$  diagrams for different states of NGC~7469. 
Data are taken from {\it RXTE} observations 80152-05-04-00 
({green} line), 10293-01-01-09 ({red} line), and 10315-01-15-10 ({blue} line).
}
\label{RXTE_spectra}
\end{figure*}


We find that the thermal model (blackbody) fits well the low energy part, while it provides an 
excess emission for $E>$~3 keV (e.g., for $A5$ 
spectrum, $\chi^2_{red}$=15.76 (161 d.o.f.), see 
the  top of Table~\ref{tab:par_asca}). As we demonstrate, the  blackbody  model points towards very large 
absorption (greater than $3\times 10^{21}$ cm$^{-2}$), particularly for the $A4$ 
spectrum, 
and moreover this model  gives unacceptable fit quality, $\chi^2$  
 for all spectra of $ASCA$ data. 
However, we must admit that the phabs*power-law model provides better fits 
than the thermal one. All {\it ASCA} data cannot be fitted by any single-component model. 
Indeed, a simple power law  produces a soft excess below 0.8 keV. These significant positive 
residuals at low energies, less than 1 keV, suggest the  presence of additional emission  components 
in the spectrum. Thus  we also test a sum of the blackbody and power-law component models. The model 
parameters of this combined model are  $N_H=3\times 10^{21}$ cm$^{-2}$, 
$kT_{bb}=1.8-2.1$ keV, and $\Gamma=2.9 - 3.2$ (see more 
in Table~\ref{tab:par_asca}). 
The best fits of the {\it ASCA} spectra has been found  using  the  
{Bulk Motion Comptonization model} ({BMC XSPEC} model, \cite{tl97})   
for which $\Gamma$ ranges from 1.6  to  1.9  for all observations (see lower part in Table~\ref{tab:par_asca}) 

We also find that a fluorescent $K_{\alpha}$ line of neutral iron is clearly detected in 
the spectrum of NGC~7469 (see Fig.~\ref{asca_interm_spectrum}). Adding a Gaussian profile to the 
absorbed BMC model improves the spectral  fits. The line centroid energy and equivalent width are $E=6.4\pm 0.01$ keV and $EW=200\pm 70$ eV, 
which is in agreement with Guainazzi's 
conclusion who apply the absorbed power-law model for {\it ASCA} spectrum, A1. 
However, iron-edge absorption is not detected in the NGC~7469 spectrum; the K$_{\alpha}$ iron emission could be reprocessed by cold matter that 
surrounds the central X-ray source (Nandra \& George 1994), located far away from the central black hole. 
We should emphasize that all these best-fit results are found using the same model for 
 all spectral  states. 

{
Previously, a number of papers (see, e.g., Koss et al. 2015 and Paggi et al. 2012) considered a possible contribution from the narrow line region (NLR)   in reference
to Seyfert 2 galaxies.   These  sources are characterized  by strong radio jets, while our source (NGC 7469) has   a low level of radio emission without any strong indication of jet-ejection events. Possibly, this fact points to  the  low/moderate
mass accretion rate in NGC 7469, 
and thus the matter outflow (or wind) 
region around the nucleus is not particularly dense as in the case of Mrk 573 (Paggi et al., 2012). Following  Paggi et al. (2012) we tested an image of this galaxy in the soft X-ray emission (see Sect.~1 as well as Figs.~\ref{imagea} and \ref{imageb}) with Swift/RXT (0.3-10 keV). Contour levels demonstrate an absence of the X-ray jet-like (elongated) structure  and show that X-ray emission is extended up to ~5 kpc (Fig.~\ref{imageb}). 

 Because  NLR emission drops rapidly towards  energies above 2 keV, our {\it RXTE} spectra (in the 3 keV$<E<$50 keV range) are not affected by it. However, we can  test the NLR effect using the  ASCA data. The estimates of the NLR contribution in Seyfert 1 galaxy NGC~7469 
based on previous $Chandra$ data (Scott et al., 2005) show that the contribution of the NLR region is low. Specifically, we assumed that this component is not variable on these timescales. While the $Chandra$ observations were carried out in 2002, we apply their main results for spectral analysis of the {\it  ASCA} data (1993, see Table 1). We included the emission lines 
and edge features into our spectral model and fitted the observed {\it ASCA} spectra (see Fig.~\ref{asca_interm_spectrum}). 
Here, the best-fit {\it ASCA} spectrum of NGC~7469 for the IS 
using the phabs*(bmc+N* Gaussian)*edge model ($\chi^2_{red}=$ 1.09  for 137 dof) is presented. The best-fit parameters 
are $\Gamma=$ 1.8$\pm$0.1, $T_s=$ 140$\pm$30 eV, and $E_{line1}=$ 6.4$\pm$0.1 keV (see more detalis in 
Tables~\ref{tab:par_asca} and ~\ref{tab:par_asca_lines}). The data are denoted by red {crosses}, while 
the spectral model presented by components is shown using  green and blue lines for BMC and Fe K$_{\alpha}$ Gaussian components, respectively. The narrow line components below 2 keV are presented by different color lines. As a result, we obtained almost the same BMC-normalization (see details in Table \ref{tab:par_asca_lines}) as  that in the model without taking account of additional emission lines (compare the results of Tables~\ref{tab:par_asca} and 
\ref{tab:par_asca_lines}). Therefore, we conclude  that the narrow lines' contribution does not affect the BMC-derived BH mass using our scaling procedure.

It is worthwhile to note that a number of papers (e.g., Koss et al., 
2015 and Paggi et al., 
2012) considered the NLR effect in detail in relation to Seyfert 2 galaxies, which have strong radio 
jets. However, our source (NGC~7469) has very low radio emission. We also inspected 
the soft X-ray image of NGC~7469 {\it Field of view} (FOV)
of a notable jet-like structure around NGC~7469. 
}

Figure~\ref{sp_ASCA_cnt} shows the best-fit model of the spectrum of NGC~7469 (top panel). 
Data are taken from $ASCA$ observations (ID=15030000) carried out on  June 26--27, 1994 when the source was in the low/hard state. The data are shown by black crosses and  
 the spectral model is displayed  by red line
 In the { bottom panel} we show  the corresponding $\Delta \chi$ versus photon energy (in keV). 
Using the $ASCA$ data  we  find that the seed temperatures $kT_s$ of the $BMC$ model  vary only slightly from 140 to 200 eV. 

We also  use the phabs*(bmc+Gaussian) model to fit  all {\it RXTE} data.
In order to  fit all of these spectra, we use neutral column $N_H$, which is fixed at 
$3\times 10^{21}$ cm$^{-2}$ (see also Behar et al, 2017; Wakker et al., 2011; Scott et al., 2005). 
In Fig.~\ref{RXTE_spectra} we demonstrate three representative $EF_E$ spectral  diagrams 
for different states of NGC~7469. Data are taken from {\it RXTE} observations 80152-05-04-00 
(green line), 10293-01-01-09 (red line), and 10315-01-15-10 ({blue} line).

%
%

\begin{table*}
 \caption{Best-fit parameters  of the $ASCA$ spectra  of NGC~7469 
 in the 0.45$-$10~keV range using the model$^\dagger$: 
phabs*(bmc+N*Gaussian)*edge. 
}
    \label{tab:par_asca_lines}
 \centering 
 \begin{tabular}{llllllll}
 \hline\hline
     & Parameter & &71028000 & 71028010 & 71028030 & 15030000 & 18931126 \\
 \hline                                             
Model &      & &       &        &       &      &     \\
      \hline
phabs      & N$_H$ ($\times 10^{21}$ cm$^{-2}$) & & 3.5$\pm$0.6 & 3.0$\pm$0.9 & 2.7$\pm$0.1  & 3.0$\pm$0.2 & 3.1$\pm$0.1 \\
bmc        & $\Gamma_{bmc}$    && 1.8$\pm$0.1 & 1.86$\pm$0.04& 1.74$\pm$0.07   & 1.81$\pm$0.06    & 1.64$\pm$0.09  \\
           & T$_{s}$   (eV)    && 140$\pm$30      & 160$\pm$20   & 200$\pm$40    & 220$\pm$30     & 210$\pm$20   \\
           & logA$$            && 0.33$\pm$0.05  & 0.35$\pm$0.04& 0.3$\pm$0.1   & 0.35$\pm$0.07  & 0.4$\pm$0.1 \\
           & N$_{bmc}^{\dagger\dagger}$ && 7.5$\pm$0.1 & 7.1$\pm$0.2 & 6.0$\pm$0.1  & 6.7$\pm$0.2  & 6.4$\pm$0.1 \\
Gaussian$^{\dagger\dagger\dagger}$   & E$_{line1}$ (keV)     & Fe K$_{\alpha}$  & 6.4$\pm$0.1 & 6.31$\pm$0.09& 6.4$\pm$0.1 & 6.42$\pm$0.04  & 6.38$\pm$0.09  \\
           & $\sigma_{line1}$ (eV) &                  & 200$\pm$10  & 150$\pm$60    & 170$\pm$90  & 110$\pm$30     & 160$\pm$30   \\
           & N$_{line1}$           &                  & 8.9$\pm$0.3 & 3.7$\pm$0.5   & 5.3$\pm$0.1  & 1.6$\pm$0.2    & 9.4$\pm$0.5 \\
           & E$_{line2}$ (keV)     & ...              & 1.5$\pm$0.1   & 1.51$\pm$0.08& 1.52$\pm$0.09 & 1.53$\pm$0.06  & 1.50$\pm$0.09  \\
           & E$_{line3}$ (keV)     & ...              & 1.13$\pm$0.09 & 1.14$\pm$0.07& 1.14$\pm$0.02 & 1.12$\pm$0.03  & 1.12$\pm$0.08  \\
           & E$_{line4}$ (keV)     & Ne X Ly $\alpha$ & 1.03$\pm$0.07 & 1.03$\pm$0.08& 1.0$\pm$0.1   & 1.04$\pm$0.05  & 1.02$\pm$0.07  \\
           & E$_{line5}$ (keV)     & ...              & 0.94$\pm$0.06 & 0.93$\pm$0.04& 0.94$\pm$0.09 & 0.94$\pm$0.08  & 0.93$\pm$0.09  \\
           & E$_{line6}$ (keV)     & Ne IX            & 0.91$\pm$0.08 & 0.90$\pm$0.08& 0.9$\pm$0.1   & 0.91$\pm$0.07  & 0.91$\pm$0.08  \\
           & E$_{line7}$ (keV)     & O XIII Ly7       & 0.86$\pm$0.09 & 0.87$\pm$0.07& 0.86$\pm$0.07 & 0.86$\pm$0.04  & 0.86$\pm$0.09  \\
           & E$_{line8}$ (keV)     & O XIII Ly6       & 0.85$\pm$0.07 & 0.85$\pm$0.09& 0.85$\pm$0.06 & 0.84$\pm$0.09  & 0.85$\pm$0.07  \\
           & E$_{line9}$ (keV)     & O XIII Ly5      & 0.84$\pm$0.08 & 0.84$\pm$0.03& 0.84$\pm$0.02 & 0.84$\pm$0.07  & 0.83$\pm$0.08  \\
           & E$_{line10}$ (keV)     & ...             & 0.8$\pm$0.1   & 0.81$\pm$0.05& 0.80$\pm$0.06 & 0.81$\pm$0.09  & 0.81$\pm$0.09  \\
           & E$_{line11}$ (keV)     & ...             & 0.79$\pm$0.05 & 0.79$\pm$0.09& 0.78$\pm$0.06 & 0.77$\pm$0.02  & 0.78$\pm$0.08  \\
           & E$_{line12}$ (keV)     & O VII           & 0.74$\pm$0.03 & 0.75$\pm$0.08& 0.74$\pm$0.05 & 0.74$\pm$0.06  & 0.75$\pm$0.09  \\
           & E$_{line13}$ (keV)     & ...             & 0.73$\pm$0.02 & 0.73$\pm$0.09& 0.73$\pm$0.02 & 0.73$\pm$0.03  & 0.73$\pm$0.08  \\
           & E$_{line14}$ (keV)     & ...             & 0.72$\pm$0.08 & 0.72$\pm$0.04& 0.72$\pm$0.08 & 0.72$\pm$0.05  & 0.72$\pm$0.09  \\
           & E$_{line15}$ (keV)     & ...             & 0.71$\pm$0.04 & 0.71$\pm$0.06& 0.70$\pm$0.09 & 0.69$\pm$0.09  & 0.71$\pm$0.07  \\
           & E$_{line16}$ (keV)     & ...             & 0.67$\pm$0.09 & 0.66$\pm$0.08& 0.67$\pm$0.02 & 0.68$\pm$0.07  & 0.68$\pm$0.09  \\
           & E$_{line17}$ (keV)     & O VIII Ly $\alpha$ & 0.65$\pm$0.07  & 0.65$\pm$0.04& 0.64$\pm$0.09 & 0.64$\pm$0.06  & 0.65$\pm$0.08  \\
           & E$_{line18}$ (keV)     & ...             & 0.61$\pm$0.08 & 0.61$\pm$0.04& 0.6$\pm$0.1 & 0.61$\pm$0.05  & 0.60$\pm$0.04  \\
           & E$_{line19}$ (keV)     & OVII            & 0.62$\pm$0.05 & 0.62$\pm$0.07& 0.62$\pm$0.09 & 0.63$\pm$0.07  & 0.62$\pm$0.08  \\
edge       & E$_{edge}$ (keV)     & Si XIV Ly${\alpha}$& 2.0$\pm$0.1 & 2.01$\pm$0.06   & 2.0$\pm$0.2 & 2.04$\pm$0.06  & 2.03$\pm$0.09  \\
           & $\tau_{max}$ ($\times 10^{-3}$) & & 7.43$\pm$0.06  & 6.1$\pm$0.1    & 6.5$\pm$0.2  & 5.1$\pm$0.3     & 6.0$\pm$0.1   \\
      \hline
  & $\chi^2$ {\footnotesize (d.o.f.)}&& 1.09 (137) & 1.01 (268) & 1.07 (116)& 1.09 (162)& 0.82 (94) \\%
      \hline
 \hline                                             
 \end{tabular}
\tablefoot{ 
$^\dagger$     Errors are given at the 90\% confidence level. 
$^{\dagger\dagger}$ Normalization parameters of BMC component 
is in units of $L^{soft}_{34}/d^2_{10}$ erg s$^{-1}$ kpc$^{-2}$, 
where $L^{soft}_{34}$ is  soft photon luminosity in units of $10^{34}$ erg s$^{-1}$, $d_{10}$ is the distance to the 
source in units of 10 kpc, while 
the Gaussian component is  in units of 
keV$^{-1}$ cm$^{-2}$ s$^{-1}$ at 1 keV and 10$^{-5}\times$ total photons cm$^{-2}$ s$^{-1}$ at 1 keV. 
$T_{s}$ is the temperature of the 
seed photon component (in keV), while  
$\Gamma_{bmc}$ is the index of the BMC component. 
$^{\dagger\dagger\dagger}$ For the rest of the Gaussian-line components, $\sigma_{line}< 100$ eV and $N_{line}< 10^{-5}$ 
total photons cm$^{-2}$ s$^{-1}$ at 1 keV.
}
 \end{table*}


\begin{figure*} 
\includegraphics[width=12cm]{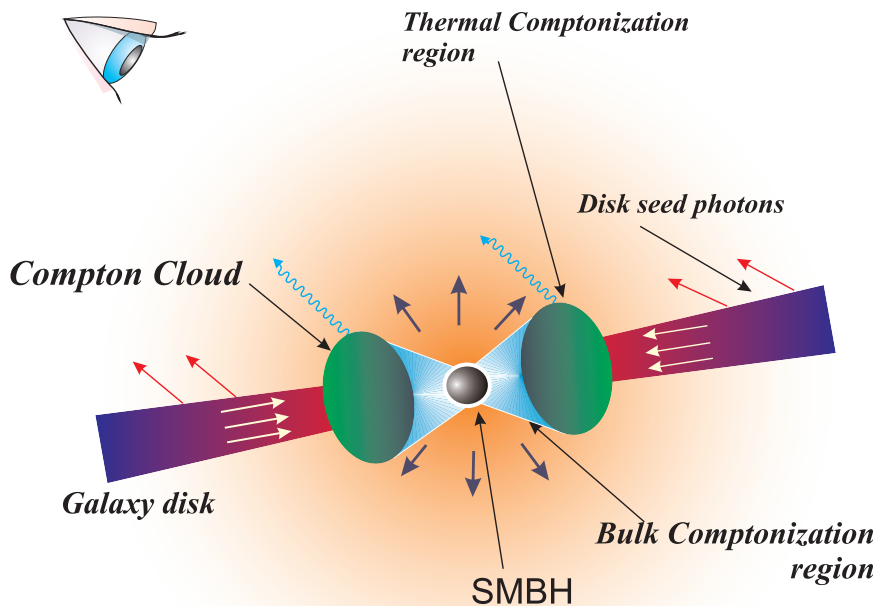}
\caption{Suggested geometry for  NGC~7469.   
Disk soft photons are upscattered (Comptonized) off  the relatively hot plasma of the transition layer (Compton cloud).
}
\label{geometry}
\end{figure*}

In Fig. \ref{geometry} we show a suggested geometry for  NGC~7469 derived using our  X-ray spectral analysis, where one can see that the soft (disk)  photons illuminate the Compton cloud (CC) surrounding a BH and finally accreting matter from the CC leads to a BH through the Bulk Comptonization region (converging flow). 

In Fig.~\ref{fragm} we present the evolution of X-ray/optical properties of NGC~7469:
the {\it RXTE}/ASM count rate (top panel), the optical V-flux (in stellar magnitudes), 
the Comptonized fraction $f$, and the BMC normalization during the 2003 outburst transition set (R2). 
In the last bottom panel, we present an evolution of the photon index $\Gamma=\alpha+1$. 
It can be seen that the soft X-ray flash (MJD 52800) is accompanied by   a normalization 
$N_{BMC}$ rise at the constant value of the photon index, $\Gamma\sim 2$.  
The high X-ray flux  (MJD 52750 -- 52800) is seen when the Comptonized fraction $f$ is low and 
 $\Gamma\sim 2$, while the subsequent transition of NGC~7469 
from the high X-ray flux phase to the low X-ray flux phase (MJD $>$ 52800) is associated with an 
 increase of the Comptonized fraction $f$ and with a decrease of the photon index ($\Gamma\sim 1.2$) 
and $N_{BMC}$. It is interesting that the optical variability (e.g., MJD 52850 -- 53000) 
is weakly related with X-ray variability. This could indicate  different origins  of the optical and X-ray emissions. 

Applying  the {\it ASCA} observations (see blue squares in Figure~\ref{saturation}), 
we find  that the spectral index $\alpha$ monotonically 
increases from 0.6 to 0.8 (or   $\Gamma$ from 1.6 to 1.8),  
when the normalization  of the BMC 
We illustrate this index versus mass accretion rate behavior  in Fig.~\ref{saturation} 
using both {\it RXTE} and {\it ASCA} observations (red triangles and blue squares, respectively).

\subsubsection{Spectral modeling for NGC~7469\label{bmc-results}}

As a result of the model selection (see Sect.~\ref{model choice}), 
we assume a model that consists 
of a sum of the Comptonization (BMC) component and the emission line (Gaussian) component to 
fit all spectral data (Tables \ref{tab:list_ASCA} and \ref{tab:list_RXTE}). 
We  briefly remind  the reader of the physical picture described  by the BMC model~(see \cite{tl97}), its basic  assumptions and  parameters. 
The BMC  Comptonization spectrum is  a sum of a part of the blackbody  directly observed  
by the Earth observer (a fraction of $1/(1+A)$) and a fraction of  the blackbody, $f=A/(1+A)$, convolved 
with the upscattering  Green's function $G(E,E_0),$ which is, in  the BMC approximation,  a broken power law 
\begin{equation}
F_{\nu}=C_N\{BB(E) +f*\int_0^{\infty}[BB(E_0)*G(E,E_0)]dE_0]\}.
\label{bmc_spectrum}
\end{equation}

It is worthwhile emphasizing that the   Green's function is  characterized by 
only one parameter, the spectral index $\alpha=\Gamma-1$. Consequently, one can see that the BMC model 
has the main parameters, $\alpha$, $A$, the seed blackbody  temperature $T_s$ , and  blackbody 
normalization, which is proportional to the seed blackbody luminosity and inversely proportional 
to $d^2$ , where d is a distance to the source. 
{ We also apply a multiplicative phabs component characterized by an equivalent hydrogen column $N_H$ 
in order to take into account  absorption by neutral material.   
The parameters of the Gaussian line component are the energy $E_{line}$ of the line centroid, 
the line width $\sigma_{line}$, and the line normalization $N_{line}$. 

One can clearly see spectral evolution  from the LHS to the IS  
in Figure~\ref{fragm}. 
The BMC model  successfully  fits  the NGC~7469 spectra for all spectral states.
 Particularly, 
the $ASCA$ spectrum obtained during the observation with ID=15030000 fit using the BMC 
model is shown in Figure~\ref{sp_ASCA_cnt}. 
In Table~\ref{tab:par_asca} 
(at the bottom), we present the results of spectral fitting  $ASCA$ data of NGC~7469 using our
 main spectral model phabs*(bmc+Gaussian). 
Using the  {\it RXTE} data, the LHS$-$IS 
transition  
is related to 
 the photon index, $\Gamma $ , which changes from 1.2 
to 2.1 when the normalization of the seed photon increases.
For the {\it RXTE} fits, we fix  the seed photon temperature at  200 eV.  
The BMC normalization, $N_{bmc}$ , varies by  a factor of thirty  in the range of 
$0.1<N_{BMC}<3.3\times L_{34}/d^2_{10}$ erg s$^{-1}$ kpc$^{-2}$, while the Comptonized (illumination) fraction changes in a wide range 
of $0.3<\log{A}<0.7$ [$f=A/(1+A)$]. 


 \begin{figure*}
 \centering
 \includegraphics[width=12cm]{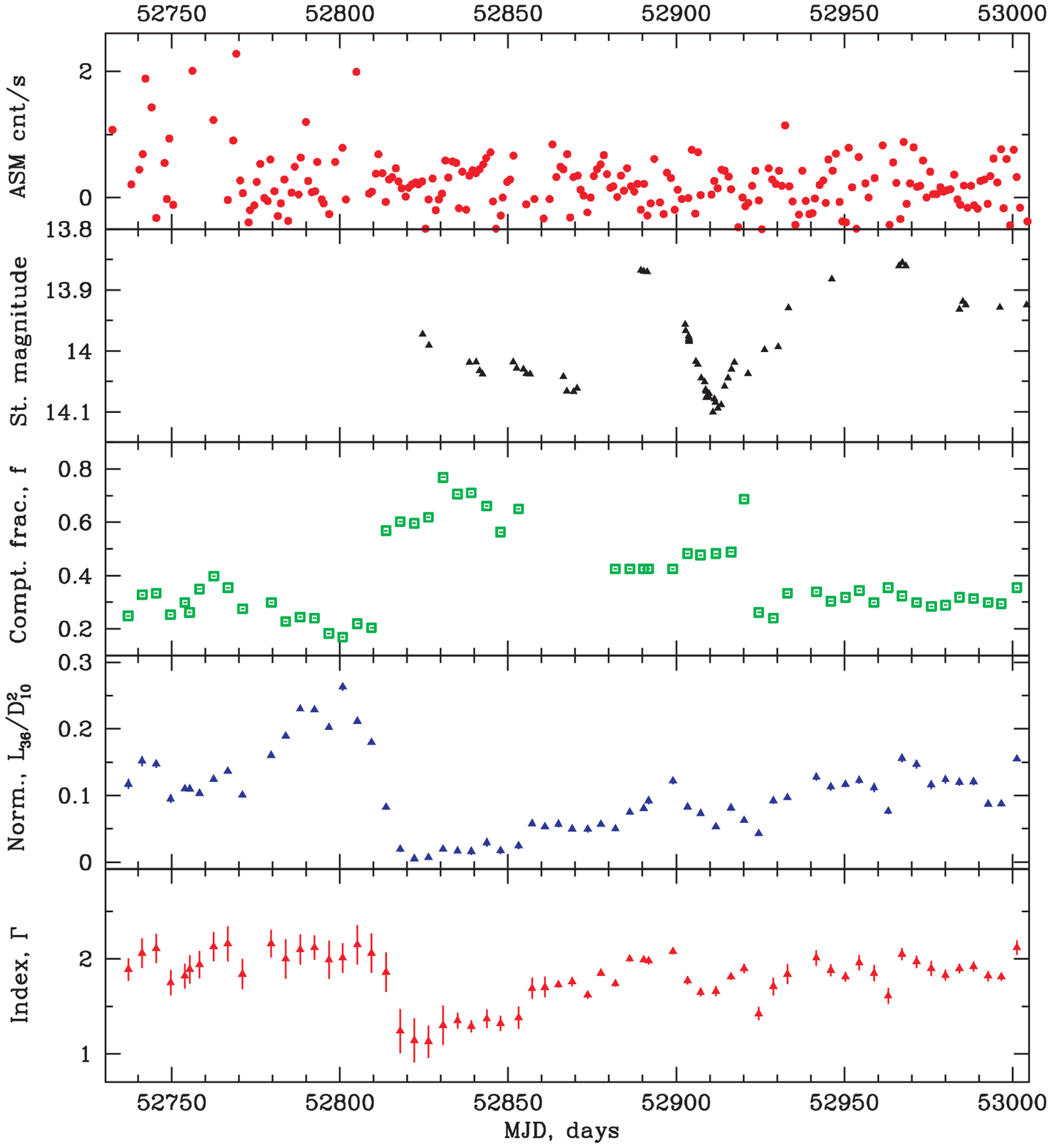}
   \caption{
{ From Top to Bottom 
:}
Evolutions of the {\it RXTE}/ASM count rate, the optical V-flux (in stellar magnitudes), the Comptonized fraction $f$, 
and the BMC normalization during the 2003 outburst transition set ($R2$). In the last bottom panel, we present an evolution 
of the photon index $\Gamma=\alpha+1$.
}
\label{fragm}
\end{figure*}

As we have already discussed  above, the spectral evolution of NGC~7469 has been previously 
investigated  using X-ray data. In particular, Guainazzi et al. (1994) 
and Rivers et al. (2013) studied the 1993 ($ASCA$) and 1996--2009 ($RXTE$) data sets, respectively 
(see also Tables \ref{tab:list_ASCA} and \ref{tab:list_RXTE}) using an additive diskbb plus power$-$law model and  a simple 
power$-$law model, 
respectively. 
These  qualitative models   describe an evolution of these spectral model parameters throughout 
state transitions during the outbursts. 

 We  also found a similar 
spectral behavior using our model and the full set of the $RXTE$ observations.  
In particular, as in the aforementioned papers by Guainazzi and Rivers, we have also revealed that 
NGC~7469 demonstrates a change of the photon index $\Gamma$ between $\sim$1.2 and 2.2
during the LHS--IS transition. In addition, we revealed that  $\Gamma$ tends to  saturate  at 2.1 
at high values of $N_{bmc}$. In other words,  we  found  that  $\Gamma$  saturates at 2.1 when  
 the mass accretion rate increases.

Our spectral model shows  a very good performance throughout
all data sets. 
{In Table~\ref{tab:par_asca} and Figs.~\ref{asca_interm_spectrum}--\ref{sp_ASCA_cnt}
 we demonstrate a good performance of the BMC model in application to the $ASCA$ and $RXTE$ data}
for which  the reduced  $\chi^2_{red}=\chi^2/N_{dof}$ 
($N_{dof}$ is the number of degree of freedom) is  less or around  1 
for  all 
observations ($0.95<\chi^2_{red}<1.12$).

%
%

 \begin{figure*}
 \centering
 \includegraphics[width=8cm]{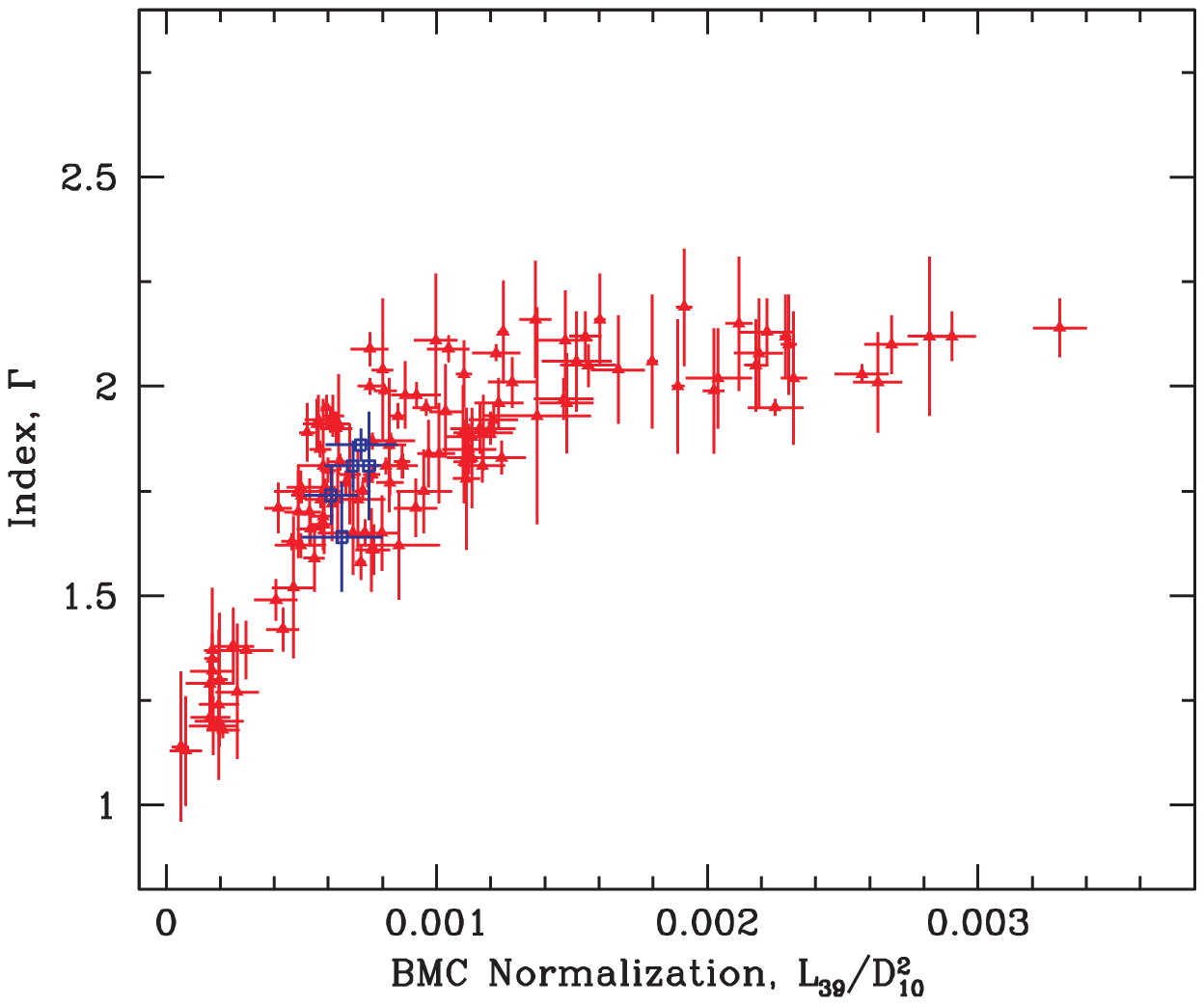}
   \caption{
Correlations of the photon index $\Gamma$ ($=\alpha+1$) 
versus  the BMC normalization, $N_{BMC}$ (proportional to mass accretion rate) in units of $L_{39}/D^2_{10}$. 
Red triangles 
and blue squares 
are related  to  $RXTE$ and $ASCA$ observations, respectively. 
}
\label{saturation}
\end{figure*}

We can also estimate a radius of the blackbody emission region.
We find 
the blackbody radius $R_{BB}$  using the relation 
$L_{BB} = 4\pi R^2_{BB}\sigma T^4_{BB}$, where $L_{BB}$ is the seed blackbody  luminosity  and 
$\sigma$ is  the Stefan-Boltzmann constant. With  a distance D to the source  of 70 Mpc, we find  the region associated with the blackbody radius of 
$R_{BB}\sim 4\times 10^{11}$ cm.  
Such an extensive  blackbody region should only be  around   a SMBH  and thus NGC~7469 is probably   a SMBH source.  
We remind the reader that 
 $R_{BB}$ is of the order of $10^6$ cm for a Galactic BH of a mass around  ten solar masses (see STS14) and 
 correspondingly $R_{BB}\sim 4\times 10^{11}$ should correspond to a characterisic size of the order of $10^6$ 
solar masses.
We also establish that 
the photon index $\Gamma$ correlates with the  BMC normalization,  $N_{BMC}$ 
(proportional to the mass accretion rate, $\dot M$) and finally saturates at higher values of $\dot M$ 
(see Figure~\ref{saturation}). 
The  index $\Gamma$
 monotonically grows from 1.2 to 2.1 with  the $\dot M$ increase
and then saturates  at $\Gamma_{sat}=2.0\pm 0.1$ for  high values of $\dot M$.

}
\section{Discussion \label{disc}}


\subsection{Saturation of the  index as a  signature of a BH  \label{constancy}}

Having applied 
our analysis to an evolution of the photon index $\Gamma$  in NGC~7469, we   find 
 the photon index, $\Gamma$ , saturates  with  the mass accretion rate, $\dot M$.
ST09 demonstrates that this index saturation is a first indication of the converging flow into a BH. 

In their early paper, \cite{tlm98} demonstrated using the equation of motion that 
the innermost part  of the accretion flow (the so-called transition layer),  shrinks  
when 
$\dot M$ increases.  
 It is worthwhile emphasizing that  for a BH,  $\Gamma$  increases and  finally saturates 
for high 
$\dot M$.  \cite{tz98}, hereafter TZ98,  semi-analytically discovered  the saturation effect  and later  \cite{LT99}, (2011), hereafter LT99 and LT11, 
confirmed this effect   making Monte Carlo simulations. 

Observations of many  Galactic BHs (GBHs) and their X-ray spectral analysis 
(see  ST09, \cite{tsei09}, \cite{ST10} and Seifina et al. (2014), hereafter TS14)
demonstrate a confrmation  of  this TZ98  prediction.
For our particular source, NGC~7469,  we also reveal  that   $\Gamma$ 
monotonically increases  from 1.2  and then   finally saturates at a value of 2.1  (see  Fig.~\ref{saturation}).  
The  index-$\dot M$ correlation found in NGC~7469  allows us to estimate a BH mass  in this 
source by scaling this correlation with those detected in a  number of GBHs and extragalactic sources  
(see details below, in Sect. 4.3).

%
%

\begin{table*}
 \caption{Parameterizations for reference and target sources.}
 \label{tab:parametrization_scal}
 \centering 
 \begin{tabular}{lcccccc}
 \hline\hline                        
  Reference source  &       $\cal A$ &     $\cal B$     &  $\cal D$  &    $x_{tr}$      & $\beta$  &  \\
      \hline
GRO~J1655--40 & 2.03$\pm$0.02 &  0.45$\pm$0.03    &  1.0 & 0.07$\pm$0.02  &   1.9$\pm$0.2  \\
Cyg~X--1      & 2.09$\pm$0.01 &  0.52$\pm$0.02    &  1.0 & 0.4$\pm$0.1    &   3.5$\pm$0.1  \\
NGC~4051      & 2.05$\pm$0.07 &  0.61$\pm$0.08    &  1.0 &   [9.54$\pm$0.2]$\times 10^{-4}$ &   0.52$\pm$0.09  \\
 \hline\hline                        
  Target source     &      $\cal A$     &    $\cal B$   &  $\cal  D$  &   $x_{tr} [\times 10^{-3}]$ & $\beta$ \\
      \hline
NGC~7469      & 2.04$\pm$0.06 & 0.62$\pm$0.03    & 1.0  &   1.25$\pm$0.04 & 0.62$\pm$0.04  \\
 \hline                                             
 \end{tabular}
 \end{table*}

\subsection{An estimate of a BH mass in NGC~7469 \label{bh_mass}}

In order to estimate BH mass, $M_{BH}$ of NGC~7469, we chose two Galactic sources 
(GRO~J1655--40, Cygnus~X--1 (see ST09)) and the extragalactic source NGC~4051 (Seifina et al. 2018, 
hereinafter SCT18), as the reference sources, 
whose BH masses and  distances  are now well established.
 
\begin{figure*}
 \centering
 \includegraphics[width=10cm]{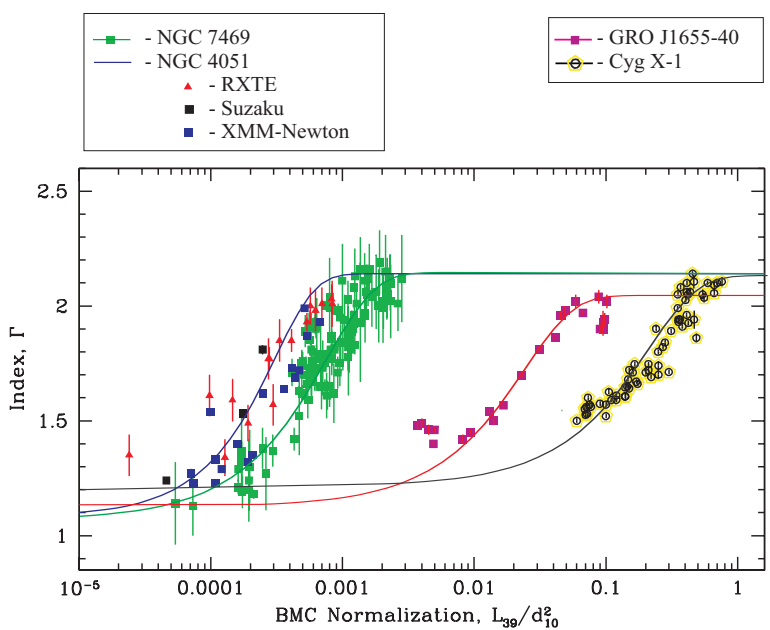}
      \caption{
Scaling of the photon index $\Gamma$ versus the normalization $N_{BMC}$ for NGC~7469 
(green line is the 
target source) as well as NGC~4051, GRO~J1655--40, and Cygnus~X--1 
(reference sources).
Red triangles stand for {\it RXTE}, 
black squares for $Suzaku,$ and blue squares for  XMM-$Newton$ data for NGC~4051. 
Pink squares and yellow-black circles correspond to {\it RXTE} data for 
GRO~J1655--40 and Cygnus~X--1, respectively.
}
\label{three_scal}
\end{figure*}

A BH mass for GRO~J1655--40 is estimated by dynamical methods. 
For a BH mass estimate 
of   NGC~7469,  
we  also use  the BMC normalizations, $N_{BMC}$ , of these reference sources.  
As a result, we scale  the index versus  $N_{BMC}$  correlations for these reference sources  with that of 
the target source   NGC 7469 (see Figure~\ref{three_scal}). 
The value of the  index saturation  is   almost the same, $\Gamma\sim2,$  for all these target and 
reference sources.   We apply the correlations found in  these four reference sources to make  a  
 comprehensive cross-check of  a BH mass estimate for NGC~7469.

%
%

{ 

The main points of our scaling method are as follows. 
Shaposhnikov \& Titarchuk (2007), hereafter ST07, used    an inverse proportionality  of a  frequency of quasi-periodic oscilations (QPO)  on BH mass in order  to estimate a BH mass. 
 ST07  
also present theoretical arguments in terms of the transition layer model
that predict the  inverse dependence of QPO frequencies
on BH mass. Therefore, as a first scaling law we write
\begin{equation}
s_\nu=\frac{\nu_r}{\nu_t}=\frac{M_t}{M_r},
\end{equation}
where subscripts $r$ and $t$  denote reference and target sources
in scaling, respectively.

The second scaling law, which we use as a basis for our analysis technique,
is a fundamental principle of inverse square intensity dependence on
the source distance, which is expressed by the equation
\begin{equation}
\frac{F_r}{F_t}=\frac{L_r}{L_t}\, \frac{d^2_t}{d^2_r} .
\end{equation}
Here $F$ stands for the source flux detected by an  observer on Earth,
$L$ is source luminosity, and $d$ is source distance. The
luminosity $L$ can be represented as
\begin{equation}
L=\frac{GM_{BH}}{R_*}\dot{M} \eta \sim \frac{GM_{BH}}{R_S} \dot{M} \eta \sim
\dot{M} \eta = M_{BH} \dot{m} \eta
,\end{equation}
where  $R_*$ is an effective radius at which energy release occurs, $\eta$ is the
efficiency of
gravitation energy conversion into radiation power, $\dot{M}$ is the
accretion rate, and
$\dot{m}$ is its dimensionless analog normalized by the Eddington
luminosity. 
Both $\dot m$ and $\eta$ are considered to be the same for two
different
sources in the same spectral state, which leads to $L_r/L_t=M_r/M_t=1/s_\nu$. 
In our analysis of energy spectra from BH sources,
we determine the normalization of seed radiation, which is supplied by an
accretion flow (disk) prior to Comptonization. The ratio of this
normalization for
two sources in the same spectral state can be written as
\begin{equation}
s_N=\frac{N_r}{N_t}=\frac{L_r}{L_t}\frac{d^2_t}{d^2_r}f_G.
\end{equation}
Here $f_G$ is a  geometry factor that comes due to the fact that the accretion
disk, which produces thermal input  for Comptonization, has a plane geometry.
Therefore, in the case of radiation coming directly from the disk, it would have
the value $f_G=(\cos i_d)_r/(\cos i_d)_t$, where $\cos i_d$ is the inclination
angle of the disk. 
When the information on the system inclination   is available, we
can use
these   $i_d $ values. 

We are now in a position to write down the final equations of our scaling
analysis.
Namely, when $s_\nu$ and $s_N$ are measured,
 the mass and the distance of the target source can be
calculated as
\begin{equation}
M_t=s_\nu M_r
\label{mass}
\end{equation}
and
\begin{equation}
d_t=d_r\biggl(\frac{s_\nu s_N}{f_G}\biggr)^{1/2}.
\label{distance}
\end{equation}
This equation allows us to estimate $s_\nu=M_t/M_t$ using values of $s_N$,  $f_G$, $d_t$ , and $d_r$.  

With Eqs. (\ref{mass}) and (\ref{distance}) in hand, the task of
BH-mass and distance measurements for a target source is reduced to
the determination of scaling coefficients $s_\nu$ and $s_N$ with
respect to the data for a reference source. This is achieved by a technique
similar to that adopted by ST07. Specifically, after scalable state transition
episodes are identified for two sources, the correlation pattern for
a reference transition is parameterized in terms of the analytical function (see also ST09)
\begin{equation}
F(x)= {\cal A} - ({\cal D}\cdot {\cal B})\ln\{\exp[(1.0 - (x/x_{tr})^{\beta})/{\cal D}] + 1\},
\label{scaling function}
\end{equation}
with $x=N_{bmc}$.

}

As can be seen from Fig.~\ref{three_scal}, the correlations of the target source (NGC~7469) 
and the reference sources  are characterized by similar shapes and index saturation levels.
Consequently, 
it allows us to make 
a reliable scaling of these correlations with that of NGC~7469.
In order to implement the   scaling technique, we introduce an analytical 
approximation  
of the $\Gamma-N_{bmc}$ correlation, 
fit by a formula (\ref{scaling function}).
 

As a result of fitting the  observed correlation  by  this function $F(x),$
we obtained a set of the best-fit parameters $\cal A$, $\cal B$, $\cal D$, $N_{tr}$, and $\beta$
(see Table \ref{tab:parametrization_scal}).  
The meaning of these parameters is  described in detail in our previous paper (Titarchuk \& Seifina (2016), hereafter TS16).
This function $F(x)$ is widely used for a description 
of the correlation of $\Gamma$ versus $N_{bmc}$ 
(\cite{sp09}, ST09, \cite{ST10}, 
STS14, \cite{ggt14}, Titarchuk \& Seifina 2016, 2017;  
Seifina et al. (2017, 2018)). 


In order to perform this BH mass determination for the target source, one should rely upon 
the same shape of the $\Gamma-N_{bmc}$ correlations for this target source and those for the reference sources.  
Accordingly, the only difference in values of $N_{bmc}$  for these three sources is   in 
   the ratio of BH mass to the squared distance, 
$M_{BH}/d^2$. 
 As one can  see from Fig.~\ref{three_scal}, 
  the index  saturation value $\cal A$  is approximately the same for the target and reference sources (see also  the second column in Table \ref{tab:parametrization_scal}). 
Evidently, for instance,  
 the shape of the correlations for  NGC~7469 (green line) and Cyg~X--1 
(black line)  
 are similar and the only difference betrween these correlations  
   is in the BMC normalization values (proportional to 
  $M_{BH}/d^2$ ratio).
  
To estimate the BH mass,  $M_t$  , of NGC~7469 (target source), one should slide 
the reference source correlation along the $N_{bmc}-$axis  to that of the target source (see Fig. \ref{three_scal}),

\begin{equation}
M_t=M_r \frac{N_t}{N_r}
\left(\frac{d_t}{d_r}
\right)^2 f_G,
\label{scaling coefficient}
\end{equation}

\noindent where $t$ and $r$ correspond to the target and reference sources, respectively;
a  geometrical factor,  
$f_G=(\cos\theta)_r/(\cos\theta)_t$, the inclination angles $\theta_r$,  
$\theta_t$ , and the distances $d_r$, $d_t$ are distances to the reference and target sources, correspondingly  (see ST09). 
One can see values of $\theta$ in  
Table \ref{tab:par_scal}  but if some of these $\theta$-values  are unavailable, then we assume that 
$f_G\sim1$. 

In Fig.~\ref{three_scal} we demonstrate   the $\Gamma-N_{bmc}$ correlation  for NGC~7469 
(green points) obtained  using  the $RXTE$ spectra  along with the correlations  for  the  
two Galactic reference sources  (GRO~J1655--40 (pink), Cygnus~X--1 (black)) and 
one extragalactic reference source NGC~4051 (blue line), which  are similar  to the 
correlations found  for the target source.  
The BH masses and distances  for each of these target-reference pairs are shown in 
Table~\ref{tab:par_scal}. 

After rearrangement,  a BH mass, $M_t$  , for NGC~7469 can be evaluated using  the formula (see TS16)
\begin{equation}
M_t=C_0 {N_t} {d_t}^2 f_G 
\label{C0 coefficient}
,\end{equation}
\noindent where 
$C_0=(1/d_r^2)(M_r/N_r)$ is the scaling coefficient for the reference source, 
BH masses $M_t$ and $M_r$ are in solar units, and $d_r$ is the distance to a particular reference source  measured in kiloparsecs. 

We use values of $M_r$, $M_t$, $d_r$, $d_t$, and $\cos (i)$ from Table~\ref{tab:par_scal} 
and then  we calculate the lowest limit of the mass, using the best-fit value of  
$N_t= (1.25\pm 0.04)\times 10^{-3}$ 
taken at the beginning of the index saturation  (see Fig. \ref{three_scal}) and measured
in units of $L_{39}/d^2_{10}$ erg s$^{-1}$ kpc$^{-2}$ (see Table \ref{tab:parametrization_scal}
 for values of the parameters of function $F(N_t)$ (Eq. 1)).
Using $d_r$, $M_r$, $N_r$ (see ST09), we found that  $C_0\sim 2.0, 1.9,$ and  $1.83$ 
for NGC~4051, GRO~J1555--40, and Cyg~X--1. 
Finally,  we obtain  $M_{7469}\ge 3\times 10^6~M_{\odot}$ 
($M_{7469}=M_t$) 
assuming $d_{7469}\sim$~70 Mpc~\citep{Behar17} and  $f_G\sim1$.
We summarize all these results  in  Table~\ref{tab:par_scal},

The obtained  BH mass estimate is in agreement with a high bolometric luminosity for NGC~7469 
and $kT_s$  value, which is in the range of 140$-$200 eV using the {\it ASCA} spectra. 
For example, Shakura \& Sunyaev, (1973) 
(see also Novikov \& Thorne, 1973) provide an effective temperature of the accretion material of 
$T_{eff}\sim T_s\propto M_{BH}^{-1/4}$.

It is also important to emphasize  
that our NGC~7469 lower mass estimate is consistent with  a SMBH mass of 
$(1 - 6) \times 10^7$ M$_{\odot}$ \citep{Peterson04,Peterson14,Shapovalova17} estimated by the 
 reverberation mapping method. 
{The derived BH mass is the lower limit estimate only, because the photon index 
versus the normalization has an uncertainty with geometrical factor $f_G=(\cos i_d)_r//(\cos i_d)_t$. Generally, the photon index 
versus the QPO frequency correlation enables us to obtain the precise BH mass (see ST09), 
but QPOs are very hard to detect in the power spectra from AGNs. 
In the Introduction  we emphasize  that a BH mass estimate of  NGC~7469 is in a fairly wide range. 
Although our BH mass estimate is only  the lower limit of that, it significantly 
constrains the range of a BH mass for NGC 7469 (see Table 6). 
}

%
%
\begin{table*}
 \caption{BH masses and distances.}
 \label{tab:par_scal}
 \centering 
 \begin{tabular}{lllllc}
 \hline\hline                        
      Source   & M$_{dyn}$ (M$_{\odot})$ & i$_{orb}$ (deg) & d (kpc)  & M$_{scal}$ (M$_{\odot}$) \\
      \hline
GRO~J1655--40  &   6.3$\pm$0.3$^{(1, 2)}$ &  70$\pm$1$^{(1, 2)}$ &  3.2$\pm$0.2$^{(3)}$    &   ... \\
Cyg~X--1       &   6.8 -- 13.3$^{(4, 5)}$ &  35$\pm$5$^{(4, 5)}$ &  2.5$\pm$0.3$^{(4, 5)}$  &  7.9$\pm$1.0 \\
NGC~4051$^{(6, 7, 8, 9, 10)}$  & ... &     ...      & $\sim$10$\times 10^3$ & $\ge 6\times 10^{5}$ \\
NGC~7469$^{(11, 12, 13)}$  & ... &     ...      & $\sim$70$\times 10^3$ & $\ge 3\times 10^{6}$ \\
 \hline                                             
 \end{tabular}
 \tablebib{
(1) Green et al. 2001; (2) Hjellming \& Rupen 1995; (3) Jonker \& Nelemans G. 2004 
(4) \cite{Herrero95}; (5) \cite{Ninkov87}; 
(6) M$^c$Hardy et al. (2004); 
(7) \cite{haba08}; 
(8) Pounds \& King (2013); 
(9) \cite{Lobban11}; 
(10) \cite{Terashima09}; 
(11) \cite{Peterson04}; 
(12) \cite{Peterson14}; and 
(13) \cite{Shapovalova17}.
}
 \end{table*}
We  derived the bolometric luminosity using  the normalization of the BMC model   and obtain one between $2\times 10^{43}$ erg/s and $6\times 10^{44}$ erg/s  (assuming isotropic radiation). 
Observations of many  Galactic BHs (GBHs) and their X-ray spectral analysis 
(see  ST09, \cite{tsei09}, \cite{ST10} and STS14, SCT18)
confirm    the TZ98  prediction that the spectral (photon) index saturates with the mass accretion rate, obviously related to the Eddington ratio.  This value is not so far from the Eddington limit for the obtained BH mass and assumed source distance (see Table 6). 
For our particular source NGC~7469,  we also reveal  that  the photon index, $\Gamma,$ 
monotonically increases  from 1.2  and then   finally saturates at a value of 2.1  (see  Figure~\ref{saturation}).  

\section{Conclusions \label{summary}} 

We found the low$-$intermediate state transitions observed in  NGC~7469 using the full 
set of $ASCA$ (1993--1994) and {\it RXTE} observations (1996--2009) and 
we demonstrated the validity of fitting  the observed spectra applying  the BMC model   
for all observations, independent of the spectral state of the source. We investigated the X-ray outburst properties of NGC~7469 and confirmed the presence of spectral state transitions during the outbursts using the index$-$normalization (or $\dot M$) correlation observed in NGC~7469, which 
were similar to those in Galactic BHs as well as to a number of extragalactic BH sources. 
In particular, we found that NGC~7469 follows the $\Gamma-\dot M$ correlation previously obtained for 
an extragalactic SMBH source, NGC~4051 (SST18), and for the Galactic BHs GRO~J1655--40 and Cyg~X--1 (ST09), for which  we take  account of the 
particular values of the $M_{BH}/d^2$ ratio (see Fig.~\ref{three_scal}).

The photon index $\Gamma$ of the NGC~7469 spectra is  in the range of $\Gamma = 1.2 - 2.1$. 
{
We tested a possible effect  of the presence of   the  narrow line region (NLR) on our results   and conclude that the NLR contribution in the spectrum is quite low in the case of NGC~7469.
}

We  also find that the peak bolometric luminosity is about $4\times 10^{44}$ erg s$^{-1}$.  
 We used the observed index-mass accretion rate correlation to estimate $M_{BH}$  in NGC~7469. 
This scaling method was  successfully applied to find  BH masses of Galactic (e.g., ST09, STS13) 
and extragalactic black holes (TS16; \citet{sp09}; \cite{ggt14}; Titarchuk \& Seifina (2017); Seifina et al. 2017, 2018).  
An application of the scaling  technique  to the X-ray data  from {\it ASCA} and {\it RXTE} observations of NGC~7469 
allows us to estimate  $M_{BH}$ for this particular source. 
We found  values of $M_{BH}\geq 3\times 10^6 M_{\odot}$.   

Furthermore, our  BH mass estimate is 
in agreement with the previous BH mass evaluations of more than $10^6$ M$_{\odot}$ derived using detailed X-ray
spectral modeling (
Peterson et al. 2004, 2014; Shapovalova et al. 2017;
Godet et al. 2012; 
Webb et al. 2012).
Combining all these estimates with  the inferred  low temperatures of the seed disk photons $kT_s$ 
, we  establish  that the compact object of NGC~7469 is likely to 
be a supermassive black hole with at least $M_{BH}> 3\times10^6 M_{\odot}$. 

\begin{acknowledgements}

We  acknowledge the interesting remarks and points of the referee.
\end{acknowledgements}


\end{document}